%% file: main.tex
\documentclass[a4paper,11pt]{article}
\pdfoutput=1 

\usepackage{jcappub} 

\usepackage[T1]{fontenc} 
\usepackage{braket}
\usepackage{natbib}
\usepackage{color}
\usepackage[dvipsnames]{xcolor}
\usepackage{bm}
\usepackage{comment}
\usepackage{dirtytalk}
\usepackage[normalem]{ulem}

\DeclareMathOperator\erf{erf}
\definecolor{burgundy}{rgb}{0.5, 0.00, 0.13}

\newcommand{\ashmeet}[1]{\textcolor{blue}{\textbf{(AS: #1)}}}

\newcommand{\eg}{e.g$.$ }
\newcommand{\Eg}{E.g$.$ }

\newcommand{\ie}{i.e$.$ }
\newcommand{\cf}{cf$.$ }

\newcommand{\wrt}{wrt$.$ }
\newcommand{\dd}{\ensuremath{\mathrm{d}}}

\newcommand{\secref}[1]{Section~\ref{sec:#1}}
\newcommand{\appref}[1]{Appendix~\ref{app:#1}}
\newcommand{\figref}[1]{Figure~\ref{fi:#1}}
\newcommand{\eqnref}[1]{Equation~\ref{eq:#1}}

\title{Toolkit for Scalar Fields in Universes with finite-dimensional Hilbert Space}

\author[a,b]{Oliver Friedrich}
\author[c]{Ashmeet Singh}
\author[d]{Olivier Doré}

\affiliation[a]{Fakultät für Physik, Ludwig-Maximilians Universität München,\\ Geschwister-Scholl-Platz 1,
80539 München, Germany}
\affiliation[b]{Kavli Institute for Cosmology, University of Cambridge,\\CB3 0HA Cambridge, United Kingdom}
\affiliation[c]{Department of Physics,
California Institute of Technology,\\
1200 E. California Blvd., Pasadena, CA 91125, USA}
\affiliation[d]{Jet Propulsion Laboratory,
California Institute of Technology,\\
4800 Oak Grove Drive, Pasadena, CA 91109, USA}

\emailAdd{oliver.friedrich@physik.lmu.de}
\emailAdd{ashmeet@caltech.edu}
\emailAdd{olivier.p.dore@jpl.nasa.gov}

\abstract{The holographic principle suggests that the Hilbert space of quantum gravity is locally finite-dimensional. Motivated by this point-of-view, and its application to the observable Universe, we introduce a set of numerical and conceptual tools to describe scalar fields with finite-dimensional Hilbert spaces, and to study their behaviour in expanding cosmological backgrounds. These tools include accurate approximations to compute the vacuum energy of a field mode $\mathbf{k}$ as a function of the dimension $d_{\mathbf{k}}$ of the mode Hilbert space, as well as a parametric model for how that dimension varies with $|\mathbf{k}|$. We show that the maximum entropy of our construction momentarily scales like the boundary area of the observable Universe for some values of the parameters of that model. And we find that the maximum entropy generally follows a sub-volume scaling as long as $d_{\mathbf{k}}$ decreases with $|\mathbf{k}|$. We also demonstrate that the vacuum energy density of the finite-dimensional field is dynamical and decays between two constant epochs in our fiducial construction. These results rely on a number of non-trivial modelling choices, but our general framework may serve as a starting point for future investigations of the impact of finite-dimensionality of Hilbert space on cosmological physics.}

\begin{document}
\maketitle
\flushbottom


\section{Introduction}
\label{sec:intro}
The holographic principle\cite{'tHooft:1993gx,Susskind:1994vu} states that the maximum entropy $S$ that can be accumulated inside a finite region of space $\mathcal{R}$ (with a sufficiently regular boundary $\delta\mathcal{R}$) equals the boundary area of that region divided by four times the Planck area \cite{Susskind1995, Bousso2002},
\begin{equation}
    S(\mathcal{R}) \leq \frac{|\delta\mathcal{R}|}{4\ell_{\mathrm{P}}^2}\ .
\end{equation}
 Since the maximum entropy that can be attained by a quantum system is proportional to the logarithm of the dimension of its Hilbert space, this can be interpreted such that the Hilbert space representing the region $\mathcal{R}$ must be finite-dimensional \cite{Parikh2005, Bao2017, Singh2019_essay}. This finite-dimensionality is a consequence of gravity: whereas quantum field theory without gravity has infinitely many degrees of freedom in any compact region of space, when we try to excite these degrees of freedom in the presence of gravity, many of the resulting states would collapse the region into a black hole. And black holes have a finite amount of entropy which scales as the area of their horizon. Therefore, any attempts to increase the region's entropy by creating further excitations would only increase the size of the resulting black hole, and hence the size of its supporting region, suggesting that the amount of entropy that can be localized in a compact region of space is finite \cite{Banks:2000fe,Bao2017, Banks2000, Fischler2000, Witten:2001kn,Dyson2002,Parikh:2004wh,Carroll_Singh_2018}. This argument based on local Hilbert space factors is oversimplified insofar as gauge theories typically do not permit spatial regions to be identified with unique Hilbert space factors \cite{Casini2014,giddings2015, donnelly+giddings2016_1, donnelly+giddings2016_2} (and references therein), and the more precise statement would be that the observables associated with a finite region of space should have support in only a finite-dimensional Hilbert space factor. This interpretation of the holographic principle asserts a local finite-dimensionality of Hilbert space, and it can be extended to the entire (observable) cosmos by noting that in an asymptotically de-Sitter Universe the causal patch of any observer has a finite extent \cite{Banks2001, Dyson2002}. Invoking observer complementarity, this means that the physics experienced by any observer in our Universe should be described by a finite-dimensional quantum theory \cite{Dyson2002, Parikh2005b}. 
 
If this reasoning is correct, then no quantum field theory based on a non-compact symmetry group (including any group with local Lorentz symmetry) can be a fundamental description of physics in our Universe \cite{Bao2017, Parikh2005} because all unitary representations of such groups live on infinite-dimensional Hilbert spaces. This also precludes conjugate variables which satisfy Heisenberg's canonical commutation relation (CCR; and its extensions to field theory) since the latter can only be realized on an infinite-dimensional Hilbert space. Motivated by the lack of finite-dimensional representations of conjugate operators satisfying the CCR, \cite{Singh2018} have used generalised Pauli operators (GPOs) as a framework to construct finite-dimensional analogs of conjugate Hermitian operators. These operators were then used by \cite{Singh2019_essay} to build a finite-dimensional version of a scalar quantum field. They also demonstrated that the zero-point energy of such fields is significantly reduced \wrt infinite-dimensional counterparts. This hints at potentially observable consequences of finite-dimensionality for quantum fields even when a fixed background spacetime is assumed.

Ultimately, notions of space and spacetime symmetries may be emergent phenomena of an underlying, purely quantum theory \cite{Dyson2002, vanRaamsdonk2010, Cao2017, CaoCarroll2018, Carroll_Singh_2018, Cotler2019, Carroll2021, Giddings:2018koz}, in which case it would not be surprising that familiar symmetry groups are not fundamental. According to \cite{Carroll_Singh2020}, quantum fields would then only be effective descriptions of emergent pointer observables which in turn result from emergent system-environment splits that maximise notions of locality, predictability and robustness against decoherence \cite[see also][for related thoughts]{Cotler2019}. It remains a challenge for this ``quantum first'' program to construct concrete models (of \eg cosmological physics) that incorporate these concepts of emergence. We think that the approach of \cite{Singh2018} and \cite{Singh2019_essay} for constructing finite-dimensional quantum fields can be a fruitful starting point for the development of such models \cite[see \eg][for different approaches]{Bao2017b, Cao2021}. 

In this paper, we revisit the framework of generalized Pauli operators to construct a finite-dimensional rendering of a scalar field and develop the following extensions, with an eye toward cosmological applications: 

\begin{itemize}
    \item[i.] We develop a finite-dimensional model of scalar field dynamics in a flat Friedmann-Lemaître-Robertson-Walker (FLRW) spacetime.
    \item[ii.] We investigate two distinct choices for the eigenvalue spacings of the finite-dimensional field operators. In our fiducial construction, we choose those spacings in a way that \emph{minimizes} finite-dimensional effects on the ground state energy of the field. We also show that the variance of the scalar field and the variance of its canonically conjugate momentum field in the ground state are equally well resolved with our choice of eigenvalue spacing. Both of these properties ensure that - in the ground state - our construction closely resembles the infinite-dimensional limit (which is also a prerequisite for the emergence of classicality in low energy physics).
    \item[iii.] In an alternative construction, we choose the eigenvalue spacing of the finite-dimensional field operators in a way that ensures an algebraic symmetry between the field and its conjugate momentum. We show that in this case the effect of finite-dimensionality on the energy eigenspectrum is drastically increased.
    \item[iv.] We derive accurate approximations for the ground state energy of the finite-dimensional harmonic oscillator as a function of frequency and the dimension of its Hilbert space. These approximations are numerically feasible for arbitrarily high dimension and agree with the exact calculation of \cite{Singh2018} to better than $3\%$ accuracy for dimensions $\geq 7$.
    \item[v.] We introduce a parametric model for how the dimension $d_{\bm{k}}$ of the Hilbert space corresponding to the co-moving mode $\bm{k}$ of our field depends on $|\bm{k}|$. While that model is likely to be overly simplistic, it allows us to qualitatively study how consistency with low energy physics can constrain its parameter space, and how different parameter values impact the behaviour of the finite-dimensional field. For example, we find that the maximum entropy attainable with our construction follows a sub-volume scaling with the size of the observable Universe as long as $d_{\bm{k}}$ is a decreasing function of $|\bm{k}|$, and that it can momentarily even display an area-scaling.
    \item[vi.] We study the equation of state of the vacuum energy density of the finite-dimensional scalar field as a function of the dimensionality parameters. For much of the allowed parameter space that energy density becomes dynamical. With our fiducial choice for the eigenvalue spacing of the conjugate field operators it is decaying between two constant epochs with an asymptotic suppression of vacuum energy by about $40\%$ . For our alternative choice of the eigenvalue spacing it is decaying indefinitely, easily reaching a $\sim 10^{-60}$ suppression compared to the infinite-dimensional calculation (with sharp UV cut-off) for some parameter values.
\end{itemize}
We have implemented the above framework within the \verb|GPUniverse| toolkit that is publicly available at \url{https://github.com/OliverFHD/GPUniverse}. Our paper is structured as follows: In \secref{finite_dim_theory}, we construct our finite-dimensional version of the scalar field in an expanding universe and derive expressions for its vacuum energy density, with the derivation of some key statements outsourced to Appendices \ref{app:minimizing_vacuum_energy} and \ref{app:Emin_estimate}. In \secref{interpretation_of_mode_emergence}, we investigate how the number of degrees of freedom in our field scale with the size of the universe, and we discuss a potential interpretation of that dynamical increase of Hilbert space dimension within the context of the work of \cite{Bao2017b}. In \secref{finite_dim_eos}, we study the equation of state of vacuum energy density of the finite-dimensional scalar field as a function of the parameters describing how the dimension of individual mode Hilbert spaces depends on the absolute value $|\bm{k}|$ of those modes (\cf point v.\ above). We also derive there a number of consistency requirements for those dimensionality parameters, and we investigate how the behaviour of vacuum energy density changes if we switch from our fiducial eigenvalue spacing of the field operators (\cf point ii.\ above) to the alternative choice (\cf point iii.). In \secref{discussion}, we discuss the assumptions and limitations of our construction as well as possible directions for future investigation. Throughout this paper we are working with natural units, \ie we put $\hbar = G = c = 1$, unless stated otherwise.

\section{A finite-dimensional scalar field and its vacuum energy density}
\label{sec:finite_dim_theory}

\subsection{Infinite-dimensional scalar field in an expanding box}
Let us first recap the conventional, infinite-dimensional construction of a real scalar field on a curved spacetime, with the action
\begin{align}
    S =&\ \frac{1}{2}\int \sqrt{-g}\dd^4x\ \left[ g^{\alpha\beta}\phi_{,\alpha}\phi_{,\beta} - m^2\phi^2 \right]\ ,
\end{align}
where $g$ is the determinant, and $g^{\alpha\beta}$ are the components of the inverse of the metric tensor. In a flat Friedmann universe, and using the conformal form of the metric
\begin{equation}
    \dd s^2 = a^2 \left( \dd \eta^2 - \dd \bm{x}^2 \right)\ ,\ [\cdot]' \equiv \frac{\partial}{\partial \eta}[\cdot]\ ,
\end{equation}
this becomes
\begin{align}
    S =&\ \frac{1}{2}\int \dd \eta \dd^3 x\ a^2\ \left[(\phi')^2 - (\bm{\nabla}\phi)^2 - m^2a^2 \phi^2 \right]\nonumber \\
    =&\ \frac{1}{2}\int \frac{\dd \eta \dd^3 k}{(2\pi)^3}\ a^2\ \left[|\phi_{\bm{k}}'|^2 - ( |\bm{k}|^2 + m^2a^2) |\phi_{\bm{k}}|^2 \right]\ ,
\end{align}
where in the second line we moved to Fourier space, with $\bm{k}$ labeling a co-moving Fourier mode. Expressing the Fourier transform of the field in terms of real and imaginary parts, $\phi_{\bm{k}} = A_{\bm{k}} + i B_{\bm{k}}$, we must have $A_{\bm{k}} = A_{-\bm{k}}$ and $B_{\bm{k}} = -B_{-\bm{k}}$ because $\phi$ is real. This allows us to define a new field
\begin{equation}
\label{eq:definition_small_q}
    q_{\bm{k}} = \sqrt{2}\ \left\lbrace \begin{matrix} A_{\bm{k}}\ \mathrm{for}\ k_1 \leq 0  \\ \\ B_{\bm{k}}\ \mathrm{for}\ k_1 > 0 \end{matrix}\right.
\end{equation}
such that the action can be re-written as \cite{PadmanabhanBook}
\begin{align}
    S =&\ \int \frac{\dd \eta \dd^3 k}{(2\pi)^3}\ \left[ \frac{a^2}{2}(q_{\bm{k}}')^2 - \frac{a^2( |\bm{k}|^2 + m^2a^2)}{2} q_{\bm{k}}^2 \right]\ .
\end{align}
This can be interpreted as an action corresponding to a collection harmonic oscillators with time dependent mass $a^2$ and time dependent frequency $\sqrt{|\bm{k}|^2 + m^2a^2}$ \cite{Mukhanov_Winitzki, PadmanabhanBook}. To make this analogy more explicit, let us restrict the field $\phi$ to a finite box of co-moving side length $L_c$, imposing periodic boundary conditions. This modifies the action to
\begin{align}
    S_{\mathrm{box}} =&\ \int \dd \eta\ \frac{1}{L_c^3}\sum_{\bm{k}}\ \left[ \frac{a^2}{2}(q_{\bm{k}}')^2 - \frac{a^2( |\bm{k}|^2 + m^2a^2)}{2} q_{\bm{k}}^2 \right]\ ,
\end{align}
where we have use the fact that $\dd^3 k \rightarrow \Delta k^3 = (2\pi/L_c)^3$ and the sum is over all $\bm{k} = (k_1, k_2, k_3)$ with $k_i \in \lbrace 2\pi n / L_c\ |\ n\in \mathbb{Z}  \rbrace$. In order to extract the Hamiltonian from that action, let us re-write it in terms of physical time $\dd t = a\dd \eta$, \ie
\begin{align}
    S_{\mathrm{box}} =&\ \int \dd t\ \frac{a^3}{L_c^3}\sum_{\bm{k}}\ \left[ \frac{1}{2}(\dot{q}_{\bm{k}})^2 - \frac{( |\bm{k}|^2/a^2 + m^2)}{2} q_{\bm{k}}^2 \right] \equiv \int \dd t\ L_{\mathrm{box}}\left( \left\lbrace q_{\bm{k}}\right\rbrace, \left\lbrace \dot{q}_{\bm{k}}\right\rbrace, t \right)\ .
\end{align}
Here the last equality serves as a definition of the Lagrangian $L_{\mathrm{box}}$ of the discretized field. It is literally the Lagrangian of a set of harmonic oscillators with masses $a^3/L_c^3$ and frequencies $\sqrt{|\bm{k}|^2/a^2 + m^2}$. The corresponding Hamiltonian is given by
\begin{equation}
\label{eq:classical_Hamiltonian}
    H_{\mathrm{box}}\left( \left\lbrace q_{\bm{k}}\right\rbrace, \left\lbrace p_{\bm{k}}\right\rbrace, t \right) = \sum_{\bm{k}}\ \left[ \frac{L_c^3}{2 a^3}\ p_{\bm{k}}^2 + \frac{a^3}{L_c^3}\frac{( |\bm{k}|^2/a^2 + m^2)}{2} q_{\bm{k}}^2 \right]\ ,
\end{equation}
where we introduced the conjugate momenta $p_{\bm{k}} = \partial L_{\mathrm{box}} / \partial \dot{q}_{\bm{k}}$ . To obtain the quantum theory of this field one would usually promote $q_{\bm{k}}$ and $p_{\bm{k}}$ to conjugate Hermitian operators satisfying the Heisenberg commutation relations
\begin{equation}
\label{eq:CCR}
    [\hat q_{\bm{k}}, \hat p_{\bm{k}'}] = i\delta_{\bm{k},\bm{k}'}
\end{equation}
such that the Hamiltonian operator of the field becomes
\begin{equation}
\label{eq:physical_Hamiltonian}
    \hat H(t) = \sum_{\bm{k}}\ \left[ \frac{L_c^3}{2 a^3}\ \hat p_{\bm{k}}^2 + \frac{a^3}{L_c^3}\frac{( |\bm{k}|^2/a^2 + m^2)}{2} \hat q_{\bm{k}}^2 \right]\ .
\end{equation}
which at any time $t$ has the minimum eigenvalue
\begin{equation}
\label{eq:physical_Emin}
    \lambda_{\min}\left[ \hat H(t) \right] = \frac{1}{a}\sum_{|\bm{k}| < k_{\max}} \frac{\sqrt{|\bm{k}|^2 + m^2a^2}}{2}\ .
\end{equation}
Here we have introduced a co-moving ultra-violet cut-off $k_{\max}$ in order to regularise this otherwise divergent expression. Such a sharp cut-off has been criticized because it breaks Lorentz symmetry \cite{Akhmedov2002, Martin2012}. There are however reasons to believe that the breaking of Lorentz symmetry is physical \cite{Gijosa2004, Bao2017, Mathur2020}, including the premise of this paper: finite-dimensionality of Hilbert space.

To obtain the vacuum energy density of the field we need to divide this eigenvalue by the physical volume of the box, \ie by
\begin{equation}
    V_{\mathrm{ph}} = L_{\mathrm{ph}}^3 \equiv (aL_c)^3\ ,
\end{equation}
where $L_{\mathrm{ph}} = aL$ is the physical box size. The energy density of the vacuum is then given by
\begin{equation}
\label{eq:correct_attempt_at_Evac}
    \epsilon_{\mathrm{vac}} = \frac{1}{a^4 L_c^3}\sum_{|\bm{k}| < k_{\max}} \frac{\sqrt{|\bm{k}|^2 + m^2a^2}}{2}\ .
\end{equation}
For a constant co-moving box size $L_c$ this seems to indicate that $\epsilon_{\mathrm{vac}} \propto a^{-4}$, which is the behaviour of a relativistic fluid, and not that of a cosmological constant. However, it is usually assumed that the scale regularising a QFT is some fix \emph{physical} scale $\Lambda_{\mathrm{UV}}$, which for the rest of this paper we will take to be equal to the Planck mass. In an expanding Universe we would then have $k_{\max} = a \Lambda_{\mathrm{UV}}$, such that \eqnref{correct_attempt_at_Evac} becomes
\begin{equation}
\label{eq:correct_attempt_at_Evac_v2}
    \epsilon_{\mathrm{vac}} = \frac{1}{a^4 L_c^3}
    \sum_{|\bm{k}| < a \Lambda_{\mathrm{UV}}} \frac{\sqrt{|\bm{k}|^2 + m^2a^2}}{2}\ .
\end{equation}
For $\Lambda_{\mathrm{UV}} \gg m$ the sum in this expression is proportional to $a^4$, so that $\epsilon_{\mathrm{vac}}$ is indeed constant. This is still somewhat curious, because a direct calculation of the vacuum pressure $\mathrm{p}_{\mathrm{vac}}$ from the vacuum stress-energy tensor yields $\mathrm{p}_{\mathrm{vac}} \approx \epsilon_{\mathrm{vac}}/3$ \cite{Akhmedov2002}, which is again the behaviour of a relativistic fluid. Note however, that the number of modes $\bm{k}$ over which we sum in \eqnref{correct_attempt_at_Evac_v2} is now itself a function of time, and that this compensates for the energy loss that a relativistic fluid would experience in an expanding universe \cite{Singh2021, Mathur2020}. This can be interpreted in terms of a modified continuity equation for the vacuum energy density \cite{Singh2021}.

We want to stress an important subtlety: Since the Hamiltonian in \eqnref{physical_Hamiltonian} is time dependent, it is not possible for the field to remain in a state of minimum energy. Instead, each of the $q_{\bm{k}}$ behaves as a driven harmonic oscillator and the expansion of the Universe will inevitably lead to particle production \cite{Mukhanov_Winitzki, PadmanabhanBook}. If the period of the oscillators around the cut-off $\Lambda_{\mathrm{UV}}$ is much smaller than the characteristic time scales over which $a$ changes, then particle production will be negligible and the quantum state will undergo adiabatic evolution, i.e., stay in the instantaneous minimum energy eigenstate to a good approximation as time evolves. We will employ this adiabaticity assumption for the remainder of this paper. The assumption is well justified in late-time cosmology because the time scales relevant for the recent cosmic expansion history are much longer than the period of any cut-off scale that is relevant to well understood particle physics.

\subsection{Finite-dimensional scalar field in an expanding box}
\label{sec:finite_field_in_box}

We now return to our premise that the dimension of the Hilbert space of the observable Universe should be finite. In this case, the dimensions of the Hilbert spaces corresponding to individual modes $\bm{k}$ also need to be finite. In the conventional infinite-dimensional setting, such as non-relativistic quantum mechanics of a single particle, classical conjugate variables $\hat{q}$ and $\hat{p}$ are promoted to Hermitian Hilbert space operators which obey the Heisenberg canonical commutation relation (CCR)
\begin{equation}
  \label{CCR}
  [ \hat{q} , \hat{p} ] = i,
\end{equation}
where we have set $\hbar = 1$.
In a quantum field theory, the field and its conjugate momentum are operator-valued functions on spacetime which obey a continuous version of the CCR, labelled by the field modes, as done in the previous section.  The Stone-von~Neumann theorem guarantees that there is an irreducible representation of the CCR, which is unique up to unitary equivalence, on any infinite-dimensional Hilbert space that is separable (i.e., that possesses a countable dense subset) \cite{Kronz2005}.
However, in this case, the theorem also implies that the operators $\hat{\phi}$ and $\hat{\pi}$ must be unbounded.
There are therefore no irreducible representations of \eqnref{CCR} on finite-dimensional Hilbert spaces, and one needs to consider a more general algebraic structure than the one imposed by Heisenberg's CCR. 

Before we switch to a finite-dimensional construction, let us define convenient, dimensionless versions of our conjugate variables as
\begin{equation}
\label{eq:qp_to_QP}
    Q_{\bm{k}} \equiv q_{\bm{k}}/L_c^2\ ,\ P_{\bm{k}} \equiv p_{\bm{k}}L_c^2\ .
\end{equation}
We would like to promote these to finite-dimensional, hermitian operators $\hat Q_{\bm{k}},\ \hat P_{\bm{k}}$ that still allow for the emergence of semi-classical physics in the infinite-dimensional limit. In order to achieve this we follow the ansatz of \cite{Singh2018, Singh2019_essay} and model $\hat Q_{\bm{k}},\ \hat P_{\bm{k}}$ in terms of generalized Pauli operators $\hat A_{\bm{k}},\ \hat B_{\bm{k}}$ (GPOs) as
\begin{equation}
\label{eq:GPO_definition}
    \hat A_{\bm{k}} = \exp\left(-i\alpha_{\bm{k}} \hat P_{\bm{k}}\right)\ ,\ \hat B_{\bm{k}} = \exp\left(i\beta_{\bm{k}} \hat Q_{\bm{k}}\right)\ ,
\end{equation}
which are defined on a Hilbert space of finite dimension $d_{\bm{k}}$ and satisfy the Weyl commutation relation \cite{weyl1950theory}
\begin{equation}
\label{eq:weylCCR}
    \hat A_{\bm{k}} \hat B_{\bm{k}} = \exp\left(\frac{-2\pi i}{d_{\bm{k}}}\right) \hat B_{\bm{k}}\hat A_{\bm{k}}\ ,
\end{equation}
and the closure property $\hat A^{d_{\bm k}}_{\bm{k}} = \mathbb{I}_{d_{\bm k}} = \hat B^{d_{\bm k}}_{\bm{k}}$, where $\mathbb{I}_{d_{\bm k}}$ is the identity operator on the Hilbert space of dimension $d_{\bm k}$. \eqnref{weylCCR} above is an exponentiated form of Heisenberg's CCR in the sense that when the real parameters, $\alpha_{\bm{k}}$ and $\beta_{\bm{k}}$ satisfy 
\begin{equation}
\label{eq:relation_alpha_beta}
\alpha_{\bm{k}} \beta_{\bm{k}} = \frac{2\pi}{d_{\bm{k}}} \: ,
\end{equation}
then \eqnref{weylCCR} is equivalent to \eqnref{CCR} in the limit $d_{\bm{k}} \rightarrow \infty$. 
The operators $\hat Q_{\bm{k}}$ and $\hat P_{\bm{k}}$ defined through Equations~\ref{eq:GPO_definition} and \ref{eq:weylCCR} \emph{do} indeed admit a unitary representation on a Hilbert space with finite dimension $d_{\bm{k}}$.
Moreover, the representation is still unique up to unitary equivalence via the Stone-von~Neumann theorem, since a finite-dimensional Hilbert space is separable. For example, let the dimension of Hilbert space be $d_{\bm{k}} = 2\ell_{\bm{k}} + 1$ for some non-negative integer $\ell_{\bm{k}}$. Then the GPOs have the following matrix representation (up to unitary equivalence)
\begin{equation}
\hat A_{\bm{k}} = \left(
\begin{array}{cccccc}
0 & 0 & 0 & \cdots & 0 & 1 \\
1 & 0 & 0 & \cdots & 0 & 0 \\
0 & 1 & 0 & \cdots & 0 & 0 \\
\vdots & \vdots & \vdots & \ddots & \vdots & \vdots \\
0 & 0 & 0 & \cdots & 0 & 0 \\
0 & 0 & 0 & \cdots & 1 & 0
\end{array} \right)_{d_{\bm{k}} \times d_{\bm{k}}} \qquad \hat B_{\bm{k}} = \left(
\begin{array}{cccc}
\exp({\frac{2\pi i }{d_{\bm{k}}}}\ell_{\bm{k}}) & 0 & \cdots & 0 \\
0 & \exp({\frac{2\pi i }{d_{\bm{k}}}}(\ell_{\bm{k}}-1)) & \cdots & 0 \\
\vdots & \vdots & \ddots & \vdots \\
0 & 0 & \cdots & \exp({\frac{-2\pi i }{d_{\bm{k}}}}\ell_{\bm{k}})
\end{array} \right)_{d_{\bm{k}} \times d_{\bm{k}}},
\end{equation}
which indeed satisfy \eqnref{weylCCR}. The construction works for even dimensions as well, but we focus on odd values to streamline the notation. It can then be shown \cite{Singh2018} that the operators $\hat Q_{\bm{k}}$ and $\hat P_{\bm{k}}$ (defined from $\hat A_{\bm{k}}$ and $\hat B_{\bm{k}}$ via \eqnref{GPO_definition}) have bounded, discrete and linearly-spaced spectra which are given by
\begin{equation}
    \mathrm{Spec}(\hat Q_{\bm{k}}) = \lbrace -\ell_{\bm{k}}\alpha_{\bm{k}}\ ,\ \dots\ ,\  \ell_{\bm{k}}\alpha_{\bm{k}}\rbrace\ \ ;\ \ \mathrm{Spec}(\hat P_{\bm{k}}) = \lbrace -\ell_{\bm{k}}\beta_{\bm{k}}\ ,\ \dots\ ,\  \ell_{\bm{k}}\beta_{\bm{k}}\rbrace\ ,
\end{equation}
\ie the eigenvalue spacings of the two operators are given by $\alpha_{\bm{k}}$ and $\beta_{\bm{k}}$ respectively. It can be shown that in the limit $d_{\bm{k}} \rightarrow \infty$ the commutator of $\hat Q_{\bm{k}}$ and $\hat P_{\bm{k}}$ indeed approaches the infinite-dimensional CCR \cite{Singh2018}.

We would now like to quantize the Hamiltonian of \eqnref{classical_Hamiltonian} in terms of these finite-dimensional operators. Taking into account the re-scaling from \eqnref{qp_to_QP}, the Hamiltonian operator becomes
\begin{align}
\label{eq:finite_dim_Hamiltonian}
    \hat H\ =\ \sum_{|\bm{k}| < a \Lambda_{\mathrm{UV}}} \left[\frac{\hat P_{\bm{k}}^2}{2a^3 L_c} + \frac{a^3 L_c\ ( |\bm{k}|^2/a^2 + m^2)}{2} \hat Q_{\bm{k}}^2\right]\ \equiv\ \sum_{|\bm{k}| < a \Lambda_{\mathrm{UV}}} \left[\frac{\hat P_{\bm{k}}^2}{2 M} + \frac{M\Omega_{\bm{k}}^2}{2} \hat Q_{\bm{k}}^2\right]\ ,
\end{align}
where we have defined $M = a^3 L_c$ and $\Omega_{\bm{k}} = \sqrt{|\bm{k}|^2/a^2 + m^2}$ to make each individual mode formally resemble a standard quantum harmonic oscillator with time dependent mass $M$ and time dependent frequency $\Omega_{\bm{k}}$. To determine the energy spectrum of this finite-dimensional constructions, we need to fix two ingredients: the dimension $d_{\bm{k}}$ of the Hilbert space of each mode $\bm{k}$, and the spacing $\alpha_{\bm{k}}$ of the eigenvalues of $\hat Q_{\bm{k}}$ (which via \eqnref{relation_alpha_beta} also fixes the eigenvalue spacing of $\hat P_{\bm{k}}$).

As a proof of concept, \cite{Singh2018} have considered the situation where $\alpha_{\bm{k}} = \beta_{\bm{k}} = (2\pi/d_{\bm{k}})^{1/2}$. We investigate the impact of that choice on the vacuum energy of our scalar field in \secref{alternative_spacing}, but for our fiducial construction, we opt for a different approach to selecting $\alpha_{\bm{k}}$ and $\beta_{\bm{k}}$. Recall that because of \eqnref{relation_alpha_beta} any choice of $\alpha_{\bm{k}}$ already fixes $\beta_{\bm{k}}$. Hence, for any given values of $d_{\bm{k}}$, $\Omega_{\bm{k}}$ and $M$, the minimum energy $E_{\min, \bm{k}}$ of the mode $\bm{k}$ only depends on $\alpha_{\bm{k}}$. Now to fix a choice of $\alpha_{\bm k}$, consider the time $t_{\bm k}$ when the mode $\bm k$ enters the sum of \eqnref{finite_dim_Hamiltonian}, that is, when $|\bm{k}| = a(t_{\bm{k}}) \Lambda_{\mathrm{UV}}$. This is the time when the mode $\bm k$ is initialized, and we are going to choose $\alpha_{\bm k}$ such that it maximises $E_{\min, \bm{k}}$ at that time,
\begin{equation}
    \frac{d}{d\alpha_{\bm{k}}} E_{\min, \bm{k}}(t_{\bm{k}}) = 0\ .
\end{equation}
As we show in \appref{minimizing_vacuum_energy}, the eigenvalue spacings that satisfy this criterion are exactly given by
\begin{equation}
\label{eq:fiducial_EV_spacings}
    \alpha_{\bm{k}} = \sqrt{\frac{2\pi}{d_{\bm{k}} M(t_{\bm{k}}) \Omega_{\bm{k}}(t_{\bm{k}})}}\ ,\ \beta_{\bm{k}} = \sqrt{\frac{2\pi M(t_{\bm{k}}) \Omega_{\bm{k}}(t_{\bm{k}})}{d_{\bm{k}} }}\ .
\end{equation}
It can be shown that finite-dimensionality can only decrease $E_{\min, \bm{k}}$ compared to its infinite-dimensional value (\cf \cite{Singh2018} or our \appref{Emin_estimate}). Hence, the above choice for $\alpha_{\bm{k}}$ and $\beta_{\bm{k}}$ minimizes finite-dimensional effects on the low energy spectrum of the Hamiltonian at the time $t_{\bm{k}}$ when the mode $\bm{k}$ is initialised. Since both $M$ and $\Omega_{\bm{k}}$ are functions of time, the Hamiltonian of each mode will eventually move away from that sweet spot. But as long as the vacuum energy of our field is dominated by UV modes, for which $t_{\bm{k}} \approx t_{\mathrm{today}}$, our fiducial construction can be considered as conservative \wrt finite-dimensional effects.

The eigenvalue spacings of \eqnref{fiducial_EV_spacings} can also be motivated from a different, but related point of view. The operators $\hat Q_{\bm{k}}$ and $\hat P_{\bm{k}}$ start to contribute to our scalar field and its conjugate momentum field at $t_{\bm{k}}$, \ie at the time when the mode $\bm{k}$ starts to enter the sum in \eqnref{finite_dim_Hamiltonian}. Let us assume that at this moment the sub-system corresponding to mode $\bm{k}$ is initialised in its instantaneous ground state which we denote by $\ket{0(\bm{k}, t_{\bm{k}})}$. We would like our construction to resemble the infinite-dimensional limit as closely as possible at that time of initialization. To achieve this, we employ a ``resolution criterion:'' we demand that the system at $t_{\bm{k}}$ should have the same resolution in \say{position}- and \say{momentum}-space, \ie
\begin{align}
\label{eq:resolution_criterion}
    \frac{\bra{0(\bm{k}, t_{\bm{k}})} \hat Q_{\bm{k}}^2 \ket{0(\bm{k}, t_{\bm{k}})}}{\alpha_{\bm{k}}^2} =&\ \frac{\bra{0(\bm{k}, t_{\bm{k}})} \hat P_{\bm{k}}^2 \ket{0(\bm{k}, t_{\bm{k}})}}{\beta_{\bm{k}}^2} \\
    \Rightarrow \frac{\bra{0(\bm{k}, t_{\bm{k}})} \hat Q_{\bm{k}}^2 \ket{0(\bm{k}, t_{\bm{k}})}}{\alpha_{\bm{k}}^4} =&\ \frac{\bra{0(\bm{k}, t_{\bm{k}})} \hat P_{\bm{k}}^2 \ket{0(\bm{k}, t_{\bm{k}})}}{(2\pi)^2}d_{\bm{k}}^2\nonumber \\
    \Rightarrow \alpha_{\bm{k}}^4 =&\ \frac{\bra{0(\bm{k}, t_{\bm{k}})} \hat Q_{\bm{k}}^2 \ket{0(\bm{k}, t_{\bm{k}})}}{\bra{0(\bm{k}, t_{\bm{k}})} \hat P_{\bm{k}}^2 \ket{0(\bm{k}, t_{\bm{k}})}}\frac{(2\pi)^2}{d_{\bm{k}}^2}\ .
\end{align}
This criterion ensures that the spread of $\ket{0(\bm{k}, t_{\bm{k}})}$ in the eigenbasis of $\hat Q_{\bm{k}}$ is equally well resolved by the eigenvalue spacing of $\hat Q_{\bm{k}}$ as the spread of $\ket{0(\bm{k}, t_{\bm{k}})}$ in the eigenbasis of $\hat P_{\bm{k}}$ by the eigenvalue spacing of $\hat P_{\bm{k}}$. If this was not case, then even a seemingly high resolution in $\hat Q_{\bm{k}}$-space could easily by identified as deviating from infinite-dimensional behaviour in $\hat P_{\bm{k}}$-space. This is also in line with the infinite-dimensional case where both conjugate variables are equally well resolved by construction, albeit infinitely well resolved.

To the best of our knowledge, in the finite-dimensional case, no closed form expressions for the quadratic expectation values of $\hat Q_{\bm{k}}$ and $\hat P_{\bm{k}}$ are available. We can however attempt to approximate them by the corresponding expectation values of an infinite-dimensional oscillator, which when combined with \eqnref{resolution_criterion}, results in
\begin{align}\label{eq:spacing_infinite_dim}
    \bra{0(\bm{k}, t_{\bm{k}})} \hat Q_{\bm{k}}^2 \ket{0(\bm{k}, t_{\bm{k}})} \approx \frac{1}{2M(t_{\bm{k}})\Omega_{\bm{k}}(t_{\bm{k}})}\ ,&\ \bra{0(\bm{k}, t_{\bm{k}})} \hat P_{\bm{k}}^2 \ket{0(\bm{k}, t_{\bm{k}})} \approx \frac{M(t_{\bm{k}})\Omega_{\bm{k}}(t_{\bm{k}})}{2}\nonumber \\
    \Rightarrow \alpha_{\bm{k}} \approx \sqrt{\frac{2\pi}{d_{\bm{k}} M(t_{\bm{k}}) \Omega_{\bm{k}}(t_{\bm{k}})}}\ ,&\ \beta_{\bm{k}} \approx \sqrt{\frac{2\pi M(t_{\bm{k}}) \Omega_{\bm{k}}(t_{\bm{k}})}{d_{\bm{k}} }}\ .
\end{align}
This is indeed equivalent to \eqnref{fiducial_EV_spacings}. We consider this as further demonstration that our construction is conservative and minimizes finite-dimensional effects.

We show in \appref{Emin_estimate} that with the above choice for $\alpha_{\bm{k}}, \beta_{\bm{k}}$, the minimum energy eigenvalue of the finite-dimensional harmonic oscillators at any time $t$ becomes
\begin{equation}
\label{eq:Emin_main_text}
    E_{\min, \bm{k}}(t) \approx \frac{\Omega_{\bm{k}}(t)}{2} \erf\left(\frac{\pi^{3/2} M(t_{\bm{k}}) \Omega_{\bm{k}}(t_{\bm{k}})}{12 M(t)\Omega_{\bm{k}}(t)} d_{\bm{k}}\right)\ ,
\end{equation}
which approximates the exact results of \cite{Singh2018}, making them more amenable for numerical implementation at high dimensions $d_{\bm{k}}$. For $M(t)\Omega_{\bm{k}}(t) \gg M(t_{\bm{k}}) \Omega_{\bm{k}}(t_{\bm{k}})$ the right hand side of \eqnref{Emin_main_text} can significantly deviate from the infinite-dimensional result $E_{\min, \bm{k}} = \Omega_{\bm{k}}(t)/2$.  The minimum eigenvalue of the total Hamiltonian is then
\begin{align}
    E_{\min}(t)\ \approx&\ \sum_{|\bm{k}| < a(t) \Lambda_{\mathrm{UV}}} \frac{\Omega_{\bm{k}}(t)}{2} \erf\left(\frac{\pi^{3/2} M(t_{\bm{k}}) \Omega_{\bm{k}}(t_{\bm{k}})}{12 M(t)\Omega_{\bm{k}}(t)} d_{\bm{k}}\right)
\end{align}
and the corresponding vacuum energy density is
\begin{align}
    \epsilon_{\mathrm{vac}}(t)\ \approx&\ \frac{1}{a(t)^3 L_c^3} \sum_{|\bm{k}| < a(t)\Lambda_{\mathrm{UV}}} \frac{\Omega_{\bm{k}}(t)}{2} \erf\left(\frac{\pi^{3/2} M(t_{\bm{k}}) \Omega_{\bm{k}}(t_{\bm{k}})}{12 M(t)\Omega_{\bm{k}}(t)} d_{\bm{k}}\right) \nonumber \\
    \approx&\ \frac{1}{a(t)^4} \underset{|\bm{k}| < a(t)\Lambda_{\mathrm{UV}}}{\int} \frac{\dd^3 \bm{k}}{(2\pi)^3}\ \frac{\sqrt{|\bm{k}|^2 + m^2 a(t)^2}}{2} \erf\left(\frac{\pi^{3/2} M(t_{\bm{k}}) \Omega_{\bm{k}}(t_{\bm{k}})}{12 M(t)\Omega_{\bm{k}}(t)} d_{\bm{k}}\right)
\end{align}
If this integral is dominated by high $k \equiv |\bm{k}| \approx a\Lambda_{\mathrm{UV}}$, and if the mass $m$ of the field is much smaller than $\Lambda_{\mathrm{UV}}$ (as must be the case for all standard model particles), then the frequency of the modes that dominate vacuum energy is given by $\Omega_{\bm{k}} \approx k/a$ and we can further approximate $\epsilon_{\mathrm{vac}}$ by
\begin{align}
\label{eq:epsilon_vac_massless}
    \epsilon_{\mathrm{vac}}(t)\ \approx&\ \frac{1}{a(t)^4} \underset{k < a(t)\Lambda_{\mathrm{UV}}}{\int} \frac{\dd k\ k^3}{(2\pi)^2}\ \erf\left(\frac{\pi^{3/2} M(t_{\bm{k}}) a(t)}{12 M(t) a(t_{\bm{k}})} d_{\bm{k}}\right) \nonumber \\
    =&\  \underset{k_{\mathrm{ph}} < \Lambda_{\mathrm{UV}}}{\int} \frac{\dd k_{\mathrm{ph}}\ k_{\mathrm{ph}}^3}{(2\pi)^2}\ \erf\left(\frac{\pi^{3/2} M(t_{a(t)k_{\mathrm{ph}}})a(t)}{12 M(t) a(t_{a(t)k_{\mathrm{ph}}})} d_{a(t)k_{\mathrm{ph}}}\right) \ ,
\end{align}
where $k_{\mathrm{ph}}$ is now physical (as opposed to co-moving) wave number. Without the error function in the integrand of this expression, this would be the standard result for vacuum energy density of a scalar field with a hard UV cut-off (\cf \eqnref{correct_attempt_at_Evac_v2}). In \secref{finite_dim_eos} we will see that this modification can significantly suppress $\epsilon_{\mathrm{vac}}(t)$ in a time-dependent way.

The final ingredient we are missing in order to evaluate the above result for $\epsilon_{\mathrm{vac}}(t)$ is the dimension $d_{\bm{k}}$ of the mode Hilbert spaces. It was motivated by \cite{Singh2019_essay} that $d_{\bm{k}} \lesssim 1/\pi k^2$ should be an upper bound for this dimension, based on the requirement that the maximum energy in each mode should be smaller than the Schwarzschild energy of the Universe. They however argue that this is a rather loose bound, and furthermore, their derivation was carried out with a choice for the eigenvalue spacings $\alpha_{\bm{k}}, \beta_{\bm{k}}$ that differs from our construction. In the following, we will use an agnostic parametrisation of the form
\begin{equation}
\label{eq:dimension_parametrized}
    d_{\bm{k}} = D\ (k/\Lambda_{\mathrm{UV}})^{n_D} + d_{\min}\ ,
\end{equation}
where we take $D > 0$ and $d_{\min} = 2$ to ensure that every mode is initialised with at least the Hilbert space of a qubit. Division by the physical cut-off $\Lambda_{\mathrm{UV}}$ has been introduced to keep $D$ a dimensionless parameter, but note that $k$ is still co-moving wave number. Note also that \eqnref{dimension_parametrized} should be interpreted as an approximation to what should actually be a function with discrete, integer values. In Sections~\ref{sec:interpretation_of_mode_emergence} and \ref{sec:finite_dim_eos} we investigate how different values of $D$ and $n_D$ impact the behaviour of our finite-dimensional field and how demanding consistency with low energy physics can constrain the $D-n_D$ space.

\subsection{Choice of IR scale}
\label{sec:size_of_cosmos}

In the previous subsections we have quantized our field in a box of finite physical side length $L_{\mathrm{ph}} = aL_c$. In the following we want to interpret this length as approximating the size of the observable Universe. This is a strong simplification since we would expect any meaningful boundary of the Universe to be spherical. To account for such a spherical geometry we would in principle have to change the way we discretized the field. Instead of a decomposition in terms of Fourier modes we would \eg need to expand the field in terms of 3D Zernicke polynomials \cite{Janssen2015}. We do not expect such a procedure to qualitatively change the results of the remainder of this paper, and we leave more rigorous investigation of field discretization to follow-up work. For now, let us simply identify $L_{\mathrm{ph}}$ with the radius of a spherically bounded Universe. 

What should we choose this radius to be at any given time? As we argued in \secref{intro}, our classical notion of spacetime is likely to emerge from an underlying quantum theory \cite{vanRaamsdonk2010, Cao2017, CaoCarroll2018, Carroll_Singh_2018, Cotler2019, Carroll2021}, and the correct choice of $L_{\mathrm{ph}}$ should be informed by that emergence map. To work out this map is far beyond the scope of this work, but we can at least guess a number of candidate scales. We could \eg choose the particle horizon
\begin{equation}
    L_p(t) = a(t)\int_{t_i}^t \frac{\dd t'}{a(t')} \equiv a(t)[\eta(t) - \eta_i]\ ,
\end{equation}
which defines the volume about which an observer can have information at time $t$. Other natural choices would be the curvature scale
\begin{equation}
    L_R = \frac{1}{\sqrt{R}}\ ,
\end{equation}
where $R$ is the Ricci scalar, or the Hubble scale
\begin{equation}
    L_H = \frac{1}{H}\ .
\end{equation}
We could also assume that the Universe has some constant co-moving size $L_c$ and that its physical size grows proportional to the scale factor, \ie
\begin{equation}
L_a = a L_c\ .
\end{equation}
As a proof of concept, and following \cite{Bao2017b, Singh2019_essay, Singh2021}, we will indeed consider that last choice, with the constant co-moving IR scale $L_c$ given by the asymptotic, co-moving particle horizon (\ie the co-moving particle horizon in the infinite future). In our dark energy dominated FLRW Universe, this (co-moving) horizon is indeed finite and about $4.5$ times as large as today's Hubble radius.

Choosing the co-moving IR scale to be constant brings with it a number of technical conveniences. For example, it results in a constant spacing $2\pi/L_c$ of the grid we used to discretize the field, such that the sum that \eg appears in \eqnref{finite_dim_Hamiltonian} is over sub-sets of the same set of modes $\bm{k}$ at any time. This especially means that our decomposition of the total Hilbert space into individual mode Hilbert spaces is well defined and constant in time. Another advantage of a constant co-moving scale $L_c$ is that it allows us to simplify our expression for vacuum energy density (\cf \eqnref{epsilon_vac_massless}) to 
\begin{align}
\label{eq:epsilon_vac_massless_v2}
    \epsilon_{\mathrm{vac}}(t)\ \approx&\  \underset{k_{\mathrm{ph}} < \Lambda_{\mathrm{UV}}}{\int} \frac{\dd k_{\mathrm{ph}}\ k_{\mathrm{ph}}^3}{(2\pi)^2}\ \erf\left(\frac{\pi^{3/2} a^2(t_{a(t)k_{\mathrm{ph}}})}{12 a^2(t)} d_{a(t)k_{\mathrm{ph}}}\right)
\end{align}
(where we have again assumed that the mass $m$ of the field is negligible). These are however just technical conveniences, and some aspects of our derivation will need to be revisited if $L_c$ changes in time.

\section{Increase in dimension and scaling of maximum entropy}
\label{sec:interpretation_of_mode_emergence}

\begin{figure}
\centering
  \includegraphics[width=\textwidth]{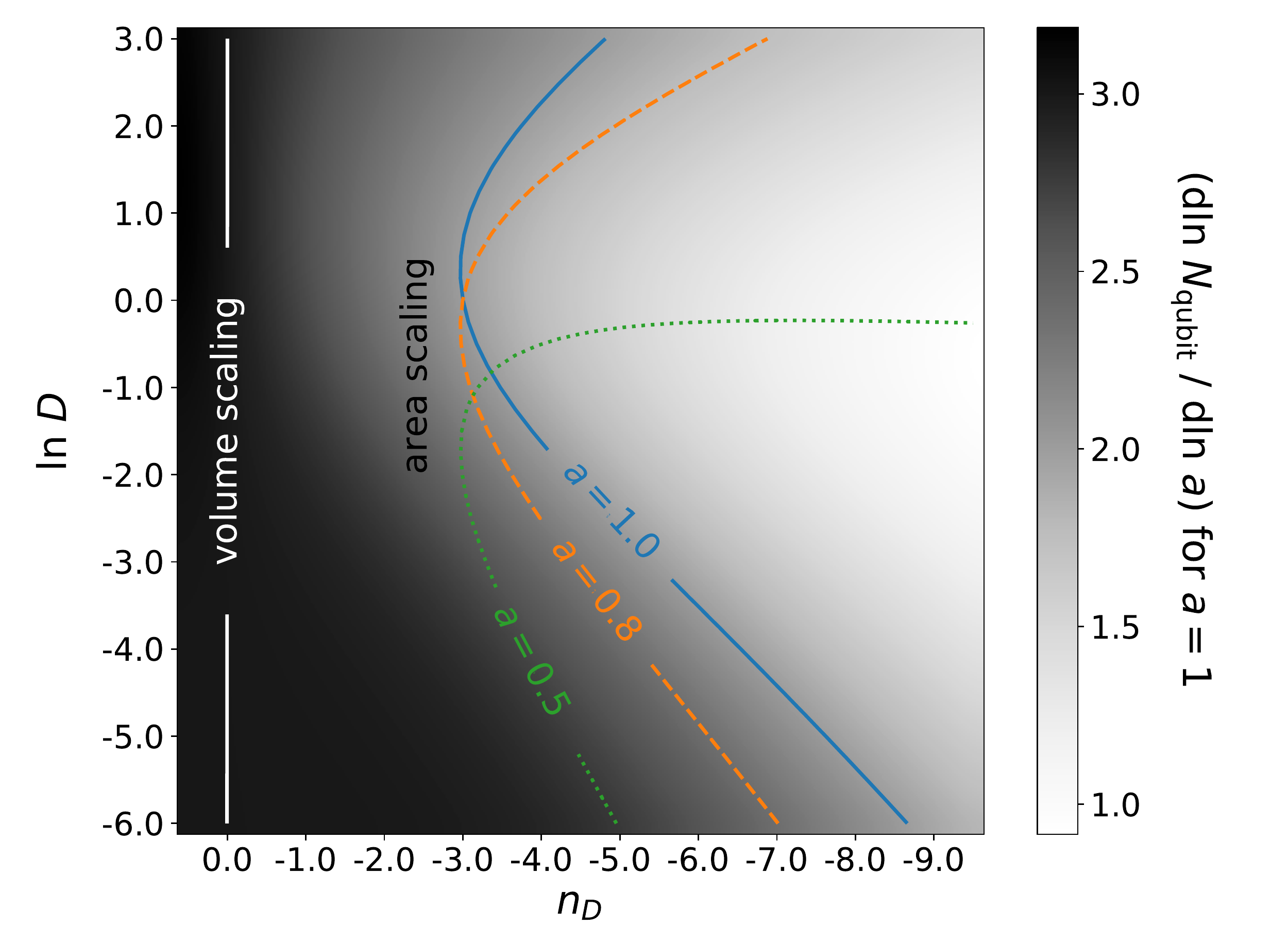}
   \caption{Color map showing how the effective number of qubits in the finite-dimensional scalar field changes with the Universe's scale factor (\cf the definition of $\gamma_{\mathrm{qubit}}$ in \eqnref{gamma_qubit}) over a range of different values for $n_D$ and $D$ and for $a = 1$. The solid blue contour indicates where $\gamma_{\mathrm{qubit}} = 2$, \ie where $N_{\mathrm{qubit}}$ follows an area scaling. For any given parameter pair $(n_D, D)$ this area scaling can only be achieved momentarily. To demonstrate this we also show the $\gamma_{\mathrm{qubit}} = 2$ contour for $a=0.8$ (dashed, orange line) and for $a=0.5$ (dotted, green line).}
  \label{fi:entropy_scaling}
\end{figure}

Recall that the Hamiltonian of our field is given by
\begin{align}
\label{eq:repeat_of_Hamiltonian}
    \hat H\ =\ \sum_{|\bm{k}| < a \Lambda_{\mathrm{UV}}} \left[\frac{\hat P_{\bm{k}}^2}{2 M} + \frac{M\Omega_{\bm{k}}^2}{2} \hat Q_{\bm{k}}^2\right]\ ,
\end{align}
In an expanding universe the upper summation limit in this expression increases with time. This could \eg be interpreted such that new modes are constantly being added to the field Hilbert space, \ie that Hilbert space dimension itself is a function of time. To avoid such a non-intuitive situation, we instead employ the view of \cite{Bao2017b}, who have modelled an expanding, constant co-moving volume $\mathcal{C}$ as a quantum circuit consisting of a number of qubits. They assume that at any time $t$ the overall Hilbert space of that circuit factorises as
\begin{equation}
    \mathcal{H} = \mathcal{H}_{\mathrm{entangled}}(t) \otimes \mathcal{H}_{\mathrm{reservoir}}(t)
\end{equation}
where $\mathcal{H}_{\mathrm{reservoir}}(t)$ consists of all of qubits in the circuit that are not entangled with any other qubit at time $t$, while each qubit in $\mathcal{H}_{\mathrm{entangled}}(t)$ is part of an entangled state. Within their framework, the entanglement of the qubits in $\mathcal{H}_{\mathrm{entangled}}(t)$ is thought to give rise to an emergent background manifold as well as to emergent, effective quantum fields on that background (\cf \cite{Cao2017, CaoCarroll2018} for a more detailed investigation of this emergence). They then model the time evolution of the total Hilbert space as a sequence of quantum gates in the circuit which entangle more and more qubits of the reservoir with qubits in $\mathcal{H}_{\mathrm{entangled}}(t)$. This leads to an increase in the dimension of $\mathcal{H}_{\mathrm{entangled}}(t)$, which \cite{Bao2017b} in turn interpret as an increase in the physical volume of $\mathcal{C}$. In an attempt to connect this picture to our construction, we could conjecture that
\begin{equation}
    \bigotimes_{|\bm{k}| < a \Lambda_{\mathrm{UV}}} \mathcal{H}_{\bm{k}}\ \subset\ \mathcal{H}_{\mathrm{entangled}}(t)
\end{equation}
where $\mathcal{H}_{\bm{k}}$ are the Hilbert spaces corresponding to the individual modes $\bm{k}$, and where we consider the left hand side to be only a subset of the left hand side in order to allow for additional degrees of freedom that constitute the background geometry. From that point of view, the modes with $|\bm{k}| \approx a \Lambda_{\mathrm{UV}}$ are not being newly created but they are simply carried over from the reservoir.

The effective number of qubits that are present in our field Hilbert space at any time $t$ is given by
\begin{equation}
    N_{\mathrm{qubit}} = \sum_{|\bm{k}| < a(t) \Lambda_{\mathrm{UV}}}\ \log_2(d_{\bm{k}})\ \approx \frac{L_c^3}{(2\pi)^3} \underset{|\bm{k}| < a(t) \Lambda_{\mathrm{UV}}}{\int}\ d^3k\ \log_2(d_{\bm{k}})\ .
\end{equation}
Note that this number is proportional to the maximum entropy that can be attained by our field. To investigate how $N_{\mathrm{qubit}}$ scales with the physical size $L_{\mathrm{ph}}(t) = a(t) L_c$, let us consider the quantity
\begin{align}
\label{eq:gamma_qubit}
    \gamma_{\mathrm{qubit}} =&\ \frac{\dd \ln N_{\mathrm{qubit}}}{\dd \ln a}\nonumber \\
    =&\ 4\pi \log_2 \left( D a^{n_D} + d_{\min} \right)\ (a \Lambda_{\mathrm{UV}})^3 \left/ \underset{|\bm{k}| < a(t) \Lambda_{\mathrm{UV}}}{\int}\ d^3k\ \log_2(d_{\bm{k}})\right.\ .
\end{align}
If $d_{\bm{k}}$ was a constant function of $|\bm{k}|$, \ie for $n_D = 0$, then $\gamma_{\mathrm{qubit}}$ would be equal to $3$ and the maximum entropy our field would obey a volume scaling. Since the holographic principle was a motivational starting point of our analysis, we are instead interested in below-volume scaling, \ie $\gamma_{\mathrm{qubit}} < 3$. As we show in \figref{entropy_scaling}, this is achieved by any $n_D < 0$. The color map in that figure shows $\gamma_{\mathrm{qubit}}$ over a range of different values for $n_D$ and $D$ and for $a = 1$, \ie in today's Universe. The extend over which we plot $n_D$ and $D$ is motivated by \secref{finite_dim_eos}, where we find that this is also the parameter range, in which the vacuum energy density of our field in today's Universe strongly deviates from a constant.

In \figref{entropy_scaling} we also show contours tracing the pairs $(n_D, D)$ for which $\gamma_{\mathrm{qubit}} = 2$, \ie for which $N_{\mathrm{qubit}}$ scales as the horizon area bounding the observable Universe. Note however, that for any pair $(n_D, D)$ such an area scaling can only be achieved momentarily. To demonstrate this, we display the area scaling contour at three different times: for $a=1$ (solid blue), $a = 0.8$ (dashed orange) and $a=0.5$ (dotted green). The fact that we cannot permanently achieve an area scaling seems to contradict the holographic principle. This problem may not be severe, because the scalar field modes $\bm{k}$ will only constitute a small part of the overall Hilbert space (which will also include spacetime degrees of freedom, \cite{Singh2019_essay, Bao2017b}) and an area scaling is only expected from the total number of degrees of freedom. We nevertheless consider this a point of concern. A potential way to enforce an area scaling would be to modify the density of modes for high $|\bm{k}|$, as was \eg investigated by \cite{Singh2019_essay} for their version of the finite-dimensional scalar field. Alternatively, it may be possible to model $d_{\bm{k}}$ with a functional form different from \eqnref{dimension_parametrized} such that an exact area scaling can be achieved at all times. As long as we are ignoring spacetime degrees of freedom we are not able to motivate either of these strategies and we choose to set aside the problem for the rest of this work.

Returning to the comparison of our construction with the quantum circuit picture of \cite{Bao2017b}, we can interpret the rate $\gamma_{\mathrm{qubit}}$ as the logarithmic rate in which entangling quantum gates are applied to the circuit as the Universe expands. This would constrain the way in which our Hamiltonian of \eqnref{repeat_of_Hamiltonian} is related to the Hamiltonian of the quantum circuit. The analogy between the two pictures however remains incomplete, because our construction attempts to model only a subset of all degrees of freedom that are present in the Universe. Also, both constructions are only approximate frameworks, and it is not obvious that a stringent mapping between the two should exist to begin with.

\section{Equation of state of finite-dimensional vacuum energy}
\label{sec:finite_dim_eos}

\subsection{Fiducial construction}
\label{sec:fiducial_construction}

\begin{figure}
\centering
  \includegraphics[width=\textwidth]{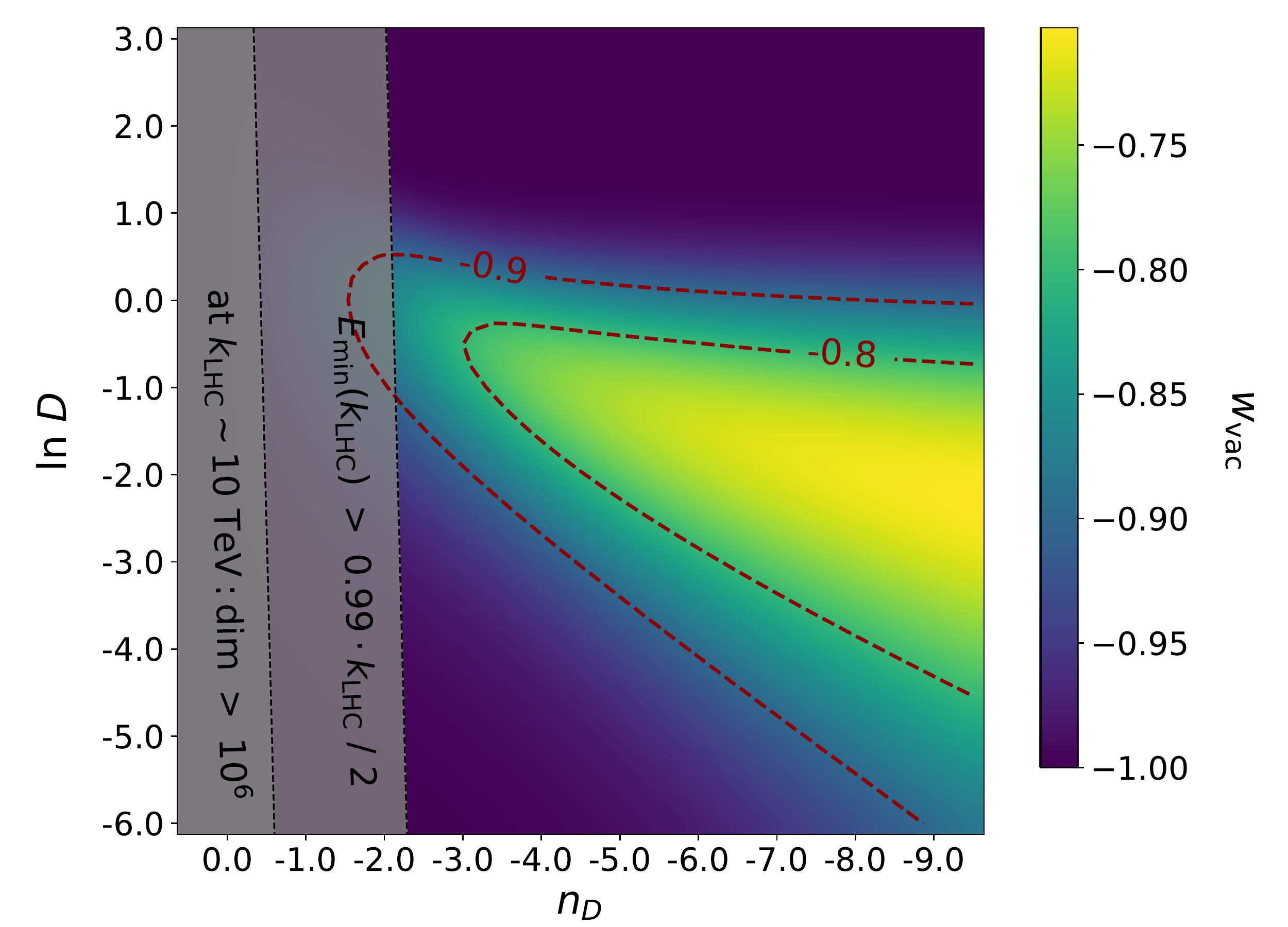}
   \caption{Color map displaying the equation-of-state parameter of the vacuum energy of our finite-dimensional scalar field as a function of the dimensionality parameters $n_D$ and $D$ introduced in \eqnref{dimension_parametrized}. The grey regions represent areas of parameter space we exclude in order to meet two consistency criteria: we require that on standard model scales the field be high-dimensional (left most bound; \cf discussion around (\ref{eq:bound_from_LHC})) and we require that on standard model scales the ground state energy of each wave mode is close to the infinite-dimensional limit (second to left most bound; \cf (\ref{eq:energy_criterion})). Both boundaries have been evaluated at a wave number $k_{\mathrm{LHC}} \sim 10$ TeV (\cf the discussion below (\ref{eq:energy_criterion}) for the impact of lower wave numbers on these bounds).}
  \label{fi:w0_as_function_of_dimension_params}
\end{figure}

\begin{figure}
\centering
  \includegraphics[width=\textwidth]{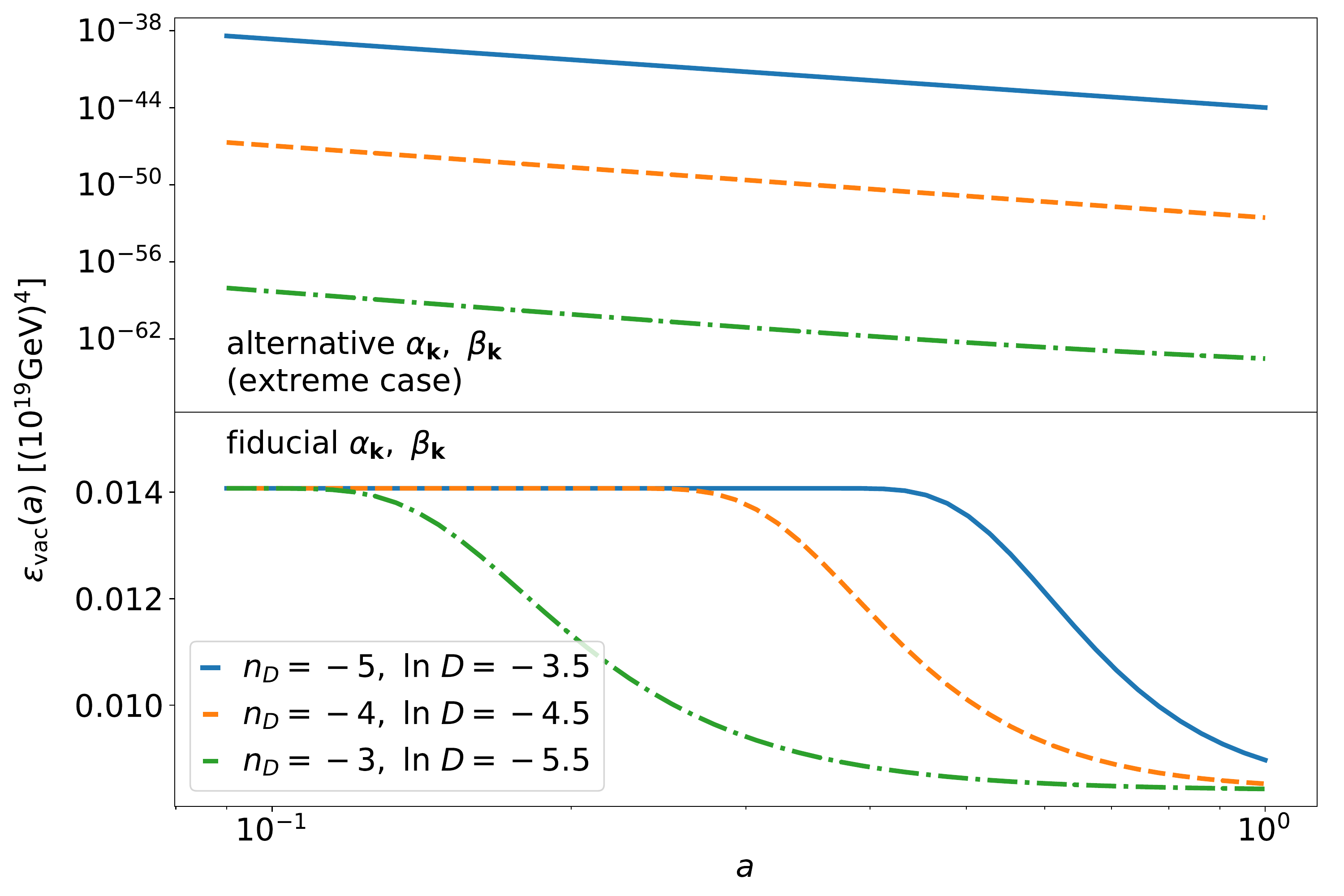}
   \caption{Evolution of vacuum energy density for different values of the dimensionality parameters $n_D$ and $D$ introduced in \eqnref{dimension_parametrized}. The bottom panel uses our fiducial construction with the eigenvalue spacings of the finite dimensional field operators given by \eqnref{fiducial_EV_spacings} while the top panel uses an extreme scenario with the eigenvalue spacings given by \eqnref{alternative_spacing} (\cf \secref{alternative_spacing}).}
  \label{fi:w_vs_a}
\end{figure}

We had argued previously, that there are several candidate scales which could act as the physical size $L_{\mathrm{ph}}$ of the Universe, and that the correct choice among those scales will depend on the mapping through which the effective notion of space emerges from an underlying purely quantum theory. Working out the details of this mapping is far beyond the scope of this work, and as a proof of concept we simply assume that the Universe has a constant co-moving size $L_c$ and that $L_{\mathrm{ph}}(t) = L_{c}\ a(t)$, \ie that the physical size of the Universe is proportional to the scale factor. As explained in \secref{size_of_cosmos}, we will furthermore choose the constant co-moving IR scale $L_c$ to be the asymptotic, co-moving particle horizon, which in our dark energy dominated FLRW Universe is about $4.5$ times as large as today's Hubble radius.

With a constant co-moving IR scale we can simplify our expression for $\epsilon_{\mathrm{vac}}$ in \eqnref{epsilon_vac_massless_v2} even further, because the scale factor $a(t_{\bm{k}})$ at the time of mode entry $t_{\bm{k}}$ can be calculated as
\begin{align}
    k_{\mathrm{ph}} a(t) =&\ k = \Lambda_{\mathrm{UV}} a(t_{\bm{k}})\nonumber \\
    \nonumber \\
    \Rightarrow \frac{k_{\mathrm{ph}}}{\Lambda_{\mathrm{UV}}} =&\ \frac{a(t_{\bm{k}})}{a(t)}\ .
\end{align}
Together with our parametrization of $d_{\bm{k}}$ from \eqnref{dimension_parametrized} this gives
\begin{align}
\label{eq:eps_vac_for_La}
    \epsilon_{\mathrm{vac}}(t)\ =&\  \underset{k_{\mathrm{ph}} \leq \Lambda_{\mathrm{UV}}}{\int} \frac{\dd k_{\mathrm{ph}}\ k_{\mathrm{ph}}^3}{(2\pi)^2}\ \erf\left(\frac{\pi^{3/2} }{12}\left[\frac{k_{\mathrm{ph}}}{\Lambda_{\mathrm{UV}}}\right]^{2} \left\lbrace D\left(\frac{a(t)k_{\mathrm{ph}}}{\Lambda_{\mathrm{UV}}}\right)^{n_D} + d_{\min}\right\rbrace\right) \ ,
\end{align}
where, as mentioned before, we take $D>0$ and $d_{\min} = 2$, such that every mode is initialised with at least the Hilbert space of a qubit. We can define an equation of state parameter $w_{\mathrm{vac}} = w_{\mathrm{vac}}(a)$ for this energy density through
\begin{align}
    \epsilon_{\mathrm{vac}} \propto &\ \exp\left( -3 \int^a \frac{\dd a'}{a'}\ (1+w) \right) \nonumber \\
    \Rightarrow w_{\mathrm{vac}} =&\ -\left(1 + \frac{1}{3} \frac{\dd \ln\epsilon_{\mathrm{vac}}}{\dd \ln a}  \right)\ .
\end{align}
We display $w_{\mathrm{vac}}$ for $a=1$ as a function of $D$ and $n_D$ in \figref{w0_as_function_of_dimension_params}. Note that the range of $D$ and $n_D$ over which we plot $w_{\mathrm{vac}}$ in that figure is the same as that of \figref{entropy_scaling}. So the region of parameter space where $w_{\mathrm{vac}}$ deviates most strongly from $-1$ seems to roughly coincide with the region in which the maximum entropy attainable with our construction deviates most from a volume-scaling.

The space of possible values for $D$ and $n_D$ should be constrained by the fact that finite-dimensional effects have not been observed on standard model scales. We have have implemented two qualitative versions of this constraint, which are displayed as grey regions in \figref{w0_as_function_of_dimension_params}. The first region results from demanding that at energies $k_{\mathrm{LHC}} \approx 10$TeV, the dimension $d_{\bm{k}}$ of the individual mode Hilbert spaces be larger than some number $N_{\mathrm{min}}$. This would require that
\begin{align}
\label{eq:bound_from_LHC}
    N_{\mathrm{min}} <&\ D (k_{\mathrm{LHC}} / \Lambda_{\mathrm{UV}})^{n_D} + d_{\min}\nonumber \\
    \Rightarrow \ln D >&\ \ln(N_{\mathrm{min}} - d_{\min}) - n_D \ln(k_{\mathrm{LHC}} / \Lambda_{\mathrm{UV}})\ .
\end{align}
It is beyond the scope of our work to determine which $N_{\mathrm{min}}$ would be sufficient to stay consistent with current experimental data (\eg regarding the Higgs boson, which so far is the only scalar field of the standard model). But for the purpose of building intuition, we implement the above bound with $N_{\mathrm{min}} = 10^6$ as the grey region to the very left of \figref{w0_as_function_of_dimension_params}. 
A second criterion we consider is the vacuum energy of modes with $k_{\mathrm{ph}}  \approx k_{\mathrm{LHC}}$. Even for large dimensions $d_{\bm{k}}$, that energy can significantly deviate from the infinite-dimensional expectation $k_{\mathrm{ph}}/2$, since it also depends on the eigenvalue spacings $\alpha_{\bm{k}}$, $\beta_{\bm{k}}$ as well as on the parameters $M$ and $\Omega_{\bm{k}}$ appearing in the Hamiltonian in \eqnref{finite_dim_Hamiltonian}. In order to ensure that finite-dimensional effects on the vacuum energy of IR scales are small we demand that
\begin{equation}
\label{eq:energy_criterion}
    \erf\left(\frac{\pi^{3/2} }{12}\left[\frac{k_{\mathrm{LHC}}}{\Lambda_{\mathrm{UV}}}\right]^{2} \left[ D \left(\frac{a(t) k_{\mathrm{LHC}}}{\Lambda_{\mathrm{UV}}}\right)^{n_D} + d_{\min}\right]\right) > x\ ,
\end{equation}
where $x$ is some number close to $1$. It is again beyond the scope of this work to determine realistic values for $x$. But as an illustration we implement the above criterion with $x=0.99$ as the second grey region to the left of \figref{w0_as_function_of_dimension_params}. Note that both Inequality~\ref{eq:bound_from_LHC} and Inequality~\ref{eq:energy_criterion} could be tightened further, because also many scales below $k_{\mathrm{LHC}}$ should be close to infinite-dimensional. \Eg, for $n_D > 0$, Inequality~\ref{eq:bound_from_LHC} would always be violated if $k_{\mathrm{LHC}}$ was replaced by an arbitrarily small scale. Similarly, for $n_D > -2$ Inequality~\ref{eq:energy_criterion} could always be violated if $k_{\mathrm{LHC}}$ was replaced by an arbitrarily small scale. It is unclear, down to which energy scales the standard model should be expected to stay accurate \cite{Jaeckel2010, DeMille2017, Budker2018}, so we do not attempt to strengthen our bounds in that way.

\eqnref{eps_vac_for_La} also allows us to consider the time dependence of vacuum energy density and we follow the evolution $\epsilon_{\mathrm{vac}}(a)$ for different values of $D$ and $n_D$ in the bottom panel of \figref{w_vs_a}. For any of the configurations $(D, n_D)$ shown there, the vacuum energy decays between two epochs of constant energy density. Correspondingly, deviations from $w=-1$ will peak at a certain time and then fall off again. This can be understood as follows: Around the upper integration limit $k_{\mathrm{ph}} \sim \Lambda_{\mathrm{UV}}$ the integrand in \eqnref{eps_vac_for_La} becomes time independent both for $a\rightarrow 0$ and $a\rightarrow \infty$ . So if the integral is dominated by that upper limit, then $\epsilon_{\mathrm{vac}}$ will become time independent in both the asymptotic past and future. We discuss this further at the end of \secref{alternative_spacing} where we also argue that \eqnref{eps_vac_for_La} is indeed dominated by $k_{\mathrm{ph}} \sim \Lambda_{\mathrm{UV}}$.

\subsection{Alternative choice of eigenvalue spacing}
\label{sec:alternative_spacing}

\begin{figure}
\centering
  \includegraphics[width=\textwidth]{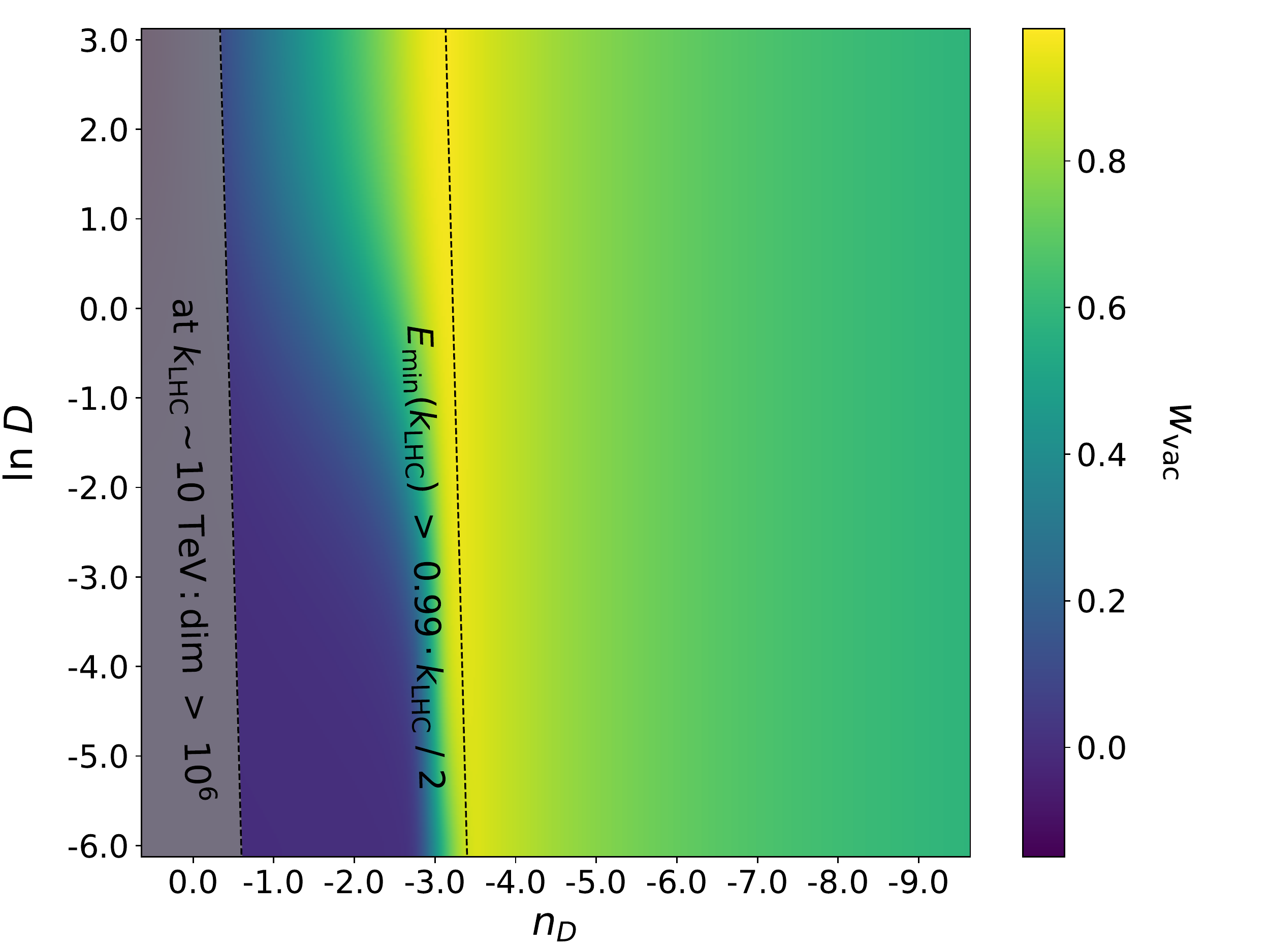}
   \caption{Same as \figref{w0_as_function_of_dimension_params} but with the alternative eigenvalue spacing of the conjugate operators as given in \eqnref{alternative_spacing}. This choice of the spacings $\alpha_{\bm{k}}$ and $\beta_{\bm{k}}$ drastically enhances finite-dimensional effects, with the equation-of-state parameter of vacuum energy reaching values up to $w_{\mathrm{vac}}=1$ . For visual purposes we now indicate the second of our consistency criteria (requiring that standard model vacuum energy be close to infinite expressions, \cf \ref{eq:energy_criterion}) only as a dashed line. Note also, that the location of that boundary has changed \wrt \figref{w0_as_function_of_dimension_params} because of the change in eigenvalue spacing.}
  \label{fi:w0_vs_dimension_params_alternative_spacing}
\end{figure}

So far we had chosen the eigenvalue spacings of our finite-dimensional conjugate operators as in \eqnref{spacing_infinite_dim}. As we have argued around that equation (see also \appref{minimizing_vacuum_energy}) this choice is minimizing finite-dimensional effects on the low-energy spectrum of Hamiltonian $\hat H_{\bm{k}}$  of each mode $\bm{k}$ at the time $t_{\bm{k}}$ when the mode emerges (\cf \secref{interpretation_of_mode_emergence} for an interpretation of the emergence process). At the same time, vacuum energy density is dominated by the UV modes for which $t_{\bm{k}}$ is close to the present time. So our construction can be considered a conservative estimate of the impact of finite-dimensionality on $\epsilon_{\mathrm{vac}}$.

We now want to complement this estimate by an alternative choice of $\alpha_{\bm{k}}$ and $\beta_{\bm{k}}$ that leads to more drastic consequences. We can do this because we take the finite-dimensional construction as more fundamental than its infinite-dimensional limit and demand only that the former approaches the latter when actually $d_{\bm{k}} \rightarrow \infty$. For our alternative scenario we follow \cite{Singh2018} in choosing
\begin{equation}
\label{eq:alternative_spacing}
    \alpha_{\bm{k}} = \sqrt{\frac{2\pi}{d_{\bm{k}}}} = \beta_{\bm{k}}\ .
\end{equation}
These values for the eigenvalue spacings are treating both conjugate operators $\hat Q_{\bm{k}}$ and $\hat P_{\bm{k}}$ equal in an algebraic sense (though we note that $\hat Q_{\bm{k}}$ and $\hat P_{\bm{k}}$ are not uniquely determined by the infinite-dimensional limit, which was our original motivation for the resolution criterion of \eqnref{resolution_criterion}).

As we show in \appref{Emin_estimate}, the vacuum energy at late time of each mode $\bm{k}$ with the new eigenvalue spacings is given by
\begin{equation}
    E_{\min, \bm{k}}(t) \approx \frac{\Omega_{\bm{k}}(t)}{2} \erf\left(\frac{\pi^{3/2}}{12 M(t)\Omega_{\bm{k}}(t)} d_{\bm{k}}\right)\ ,
\end{equation}
where as before $M = a^3 L_c$ and $\Omega_{\bm{k}} = \sqrt{|\bm{k}|^2/a^2 + m^2} \approx |\bm{k}|/a$. Hence, the vacuum energy density now becomes
\begin{align}
\label{eq:eps_vac_alternative_spacings}
    \epsilon_{\mathrm{vac}}(t)\ =&\  \underset{k_{\mathrm{ph}} \leq \Lambda_{\mathrm{UV}}}{\int} \frac{\dd k_{\mathrm{ph}}\ k_{\mathrm{ph}}^3}{(2\pi)^2}\ \erf\left(\frac{\pi^{3/2} }{12 a(t)^3\ k_{\mathrm{ph}} L_c} \left[D \left(\frac{a(t) k_{\mathrm{ph}}}{\Lambda_{\mathrm{UV}}}\right)^{n_D} + d_{\min}\right]\right) \ .
\end{align}
\begin{figure}
\centering
  \includegraphics[width=\textwidth]{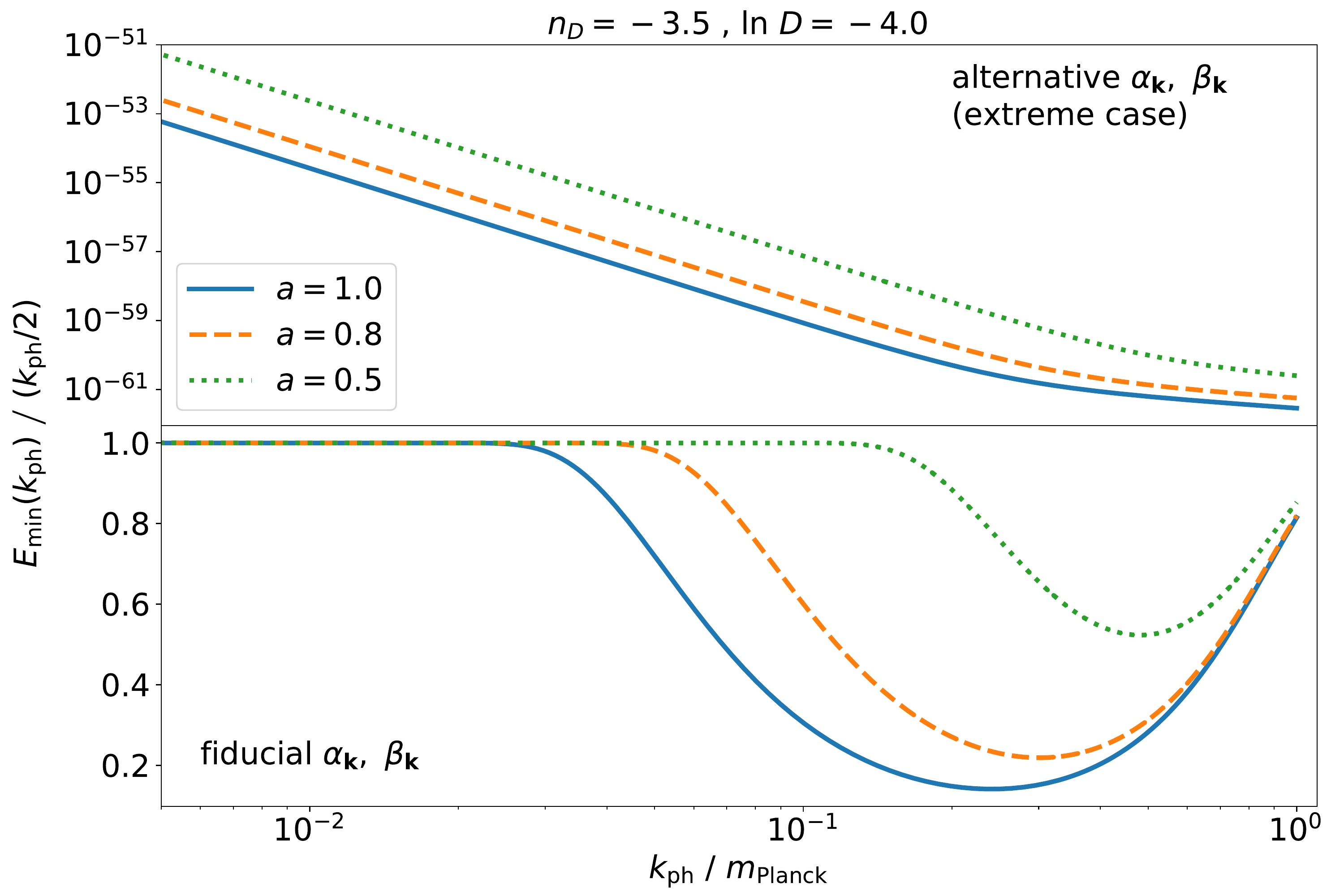}
   \caption{Showing how the error functions appearing in the integrands of \eqnref{eps_vac_for_La} (lower panel) and \eqnref{eps_vac_alternative_spacings} (upper panel) vary as a function of physical wave number $k_{\mathrm{ph}}$ when choosing the dimensionality parameters to be $(y, n_D) = (-4, -3.5)$ and for a number of different scale factors $a$. These functions encode by how much the vacuum energy of the co-moving mode $k = a k_{\mathrm{ph}}$ is reduced with respect to the infinite-dimensional expectation of $E_{\min} \approx k_{\mathrm{ph}}/2$ in our two finite-dimensional scenarios.}
  \label{fi:integrands}
\end{figure}
The top panel of \figref{w_vs_a} displays $\epsilon_{\mathrm{vac}}$ as a function of the scale factor $a$ for the same values of $D$ and $n_D$ as we had previously considered for our fiducial construction. And \figref{w0_vs_dimension_params_alternative_spacing} shows the equation of state parameter corresponding to the alternative expression for $\epsilon_{\mathrm{vac}}$ as a function of $D$ and $n_D$. The behaviour of vacuum energy is now radically altered compared to our fiducial construction (\cf \figref{w0_as_function_of_dimension_params}). Now, in most parts of parameter space, $\epsilon_{\mathrm{vac}}$ does not act as a dark energy at all and instead rapidly decays with $w_{\mathrm{vac}} \geq 0$.

To understand this strongly different behaviour, let us compare the error functions appearing in the integrands of \eqnref{eps_vac_for_La} and \eqnref{eps_vac_alternative_spacings}. As functions of physical wave number $k_{\mathrm{ph}}$ they encode the amount by which the vacuum energy of the co-moving mode $k = a k_{\mathrm{ph}}$ is reduced with respect to the infinite-dimensional expectation of $E_{\min} \approx k_{\mathrm{ph}}/2$ in our two scenarios. In our alternative scenario (\ie the one of \eqnref{eps_vac_alternative_spacings}), the error function can be approximated in terms of a piece wise power law as
\begin{equation}
\label{eq:alternative_asymtotics}
    \erf\left(\frac{\pi^{3/2} \left[D \left(\frac{a k_{\mathrm{ph}}}{\Lambda_{\mathrm{UV}}}\right)^{n_D} + d_{\min}\right]}{12 a^3 k_{\mathrm{ph}} L_c}\right) \propto \left\lbrace \begin{array}{ll} 1 &\ \ \mathrm{for}\ D \left(\frac{a k_{\mathrm{ph}}}{\Lambda_{\mathrm{UV}}}\right)^{n_D} \gg k_{\mathrm{ph}} L_c \\ \\ (k_{\mathrm{ph}})^{n_D-1} &\ \ \mathrm{for}\ k_{\mathrm{ph}} L_c \gg D \left(\frac{a k_{\mathrm{ph}}}{\Lambda_{\mathrm{UV}}}\right)^{n_D} \gg d_{\min} \\ \\ (k_{\mathrm{ph}})^{-1} &\ \ \mathrm{for}\ d_{\min} \gg D \left(\frac{a k_{\mathrm{ph}}}{\Lambda_{\mathrm{UV}}}\right)^{n_D} \end{array} \right. \ .
\end{equation}
For $n_D = -3.5$, $y = -4.0$ (a point in parameter space which results in a particularly high $w$ in our alternative construction) and at a scale factor of $a=1$ we demonstrate this behaviour in \appref{erf_asymptotics} (\cf \figref{Integrant_alternative_case_asymptotes} there). We also display the above error function for a number of different scale factors ($a = 0.5\ ,\ 0.8\ ,\ 1.0$) and on a reduced $k_{\mathrm{ph}}$-range in the upper panel of \figref{integrands}. For negative values of $n_D$ these scalings imply that deviations from the infinite-dimensional vacuum energy are monotonically increasing with $k_{\mathrm{ph}}$ (\ie the error function is monotonically decreasing, \cf \figref{integrands}) when choosing our alternative spacings for the eigenvalues of the conjugate operators. The wave number above which the error function starts to noticeably deviate from $1$ is approximately given by
\begin{equation}
    k_{\mathrm{ph}}^* \approx \left(\frac{D a^{n_D}}{(\Lambda_{\mathrm{UV}})^{n_D}L_c}\right)^{\frac{1}{1-n_D}}\ .
\end{equation}
This can be seen as a threshold above which modes start to contribute significantly less to the vacuum energy density of the field than they would in the infinite-dimensional case. For negative $n_D$ that threshold is decreasing as $a(t)$ increases, which explains the strongly decaying behaviour of vacuum energy density for the alternative eigenvalue spacings of \eqnref{alternative_spacing}.

The error function of our fiducial construction behaves quite differently. Because of an additional factor of $k_{\mathrm{ph}}^3$ in its argument, it follows a piece wise scaling of
\begin{equation}
    \erf\left(\frac{\pi^{\frac{3}{2}} k_{\mathrm{ph}}^2 \left[D \left(\frac{a k_{\mathrm{ph}}}{\Lambda_{\mathrm{UV}}}\right)^{n_D} + d_{\min}\right]}{12 \Lambda_{\mathrm{UV}}^2}\right) \propto \left\lbrace \begin{array}{ll} 1 & \mathrm{for}\ \left(\frac{\Lambda_{\mathrm{UV}}}{k_{\mathrm{ph}}}\right)^{2-n_D} \gg D a^{n_D} \\ \\ (k_{\mathrm{ph}})^{n_D+2} & \mathrm{for}\ \left(\frac{k_{\mathrm{ph}}}{\Lambda_{\mathrm{UV}}}\right)^{2} \gg D \left(\frac{a k_{\mathrm{ph}}}{\Lambda_{\mathrm{UV}}}\right)^{n_D} \gg d_{\min} \\ \\ \erf\left(\frac{\pi^{\frac{3}{2}} d_{\min}}{12 \Lambda_{\mathrm{UV}}^2} k_{\mathrm{ph}}^2\right) & \mathrm{for}\ d_{\min} \gg D \left(\frac{a k_{\mathrm{ph}}}{\Lambda_{\mathrm{UV}}}\right)^{n_D} \end{array} \right. .
\end{equation}
At least for $n_D < -2$ this implies a non-monotonic behaviour: deviations from the infinite-dimensional limit first increase with $k_{\mathrm{ph}}$ and then decrease again (\ie the error function decreases and then increases again, \cf lower panel of \figref{integrands}). And for any $n_D$ the error function at $k_{\mathrm{ph}} \sim \Lambda_{\mathrm{UV}}$ in \eqnref{eps_vac_for_La} will behave as
\begin{equation}
    \sim\ \erf\left(\frac{\pi^{\frac{3}{2}} \left[D a^{n_D} + d_{\min}\right]}{12}\right)\ ,\nonumber
\end{equation}
which is always $\sim \mathcal{O}(1)$. As a consequence, the integral in \eqnref{eps_vac_for_La} is always dominated by the UV. The resulting vacuum energy is then transitioning between two dark-energy-like regimes, as discussed at the end of \secref{fiducial_construction}: a regime when the error function equals $1$ over almost the entire integration range (which happens for $a \rightarrow 0$ as long as $n_D$ is negative) and a regime where the UV part of the integrand has approached the limit $\sim \erf(\pi^{\frac{3}{2}} d_{\min} / 12)$ (which happens for $a \rightarrow \infty$ as long as $n_D$ is negative).

The stark difference between the behaviour of vacuum energy in our fiducial and in our alternative construction highlights, that effects of finite-dimensionality are highly sensitive to the details of how quantum field theory emerges from a finite-dimensional quantum theory. Neither the fiducial construction (which minimizes the impact of finite-dimensionality) nor the alternative construction (which attempts to treat both conjugate variables as algebraically equal) are likely to fully capture this emergence. In the following section we summarize the assumptions and limitations behind both scenarios and give an outlook on potential extensions and improvements.

\section{Summary of assumptions and discussion}
\label{sec:discussion}

Starting from the point of view that the overall Hilbert space of the (observable) Universe should be finite, we have extended the framework of \cite{Singh2018, Singh2019_essay} for describing scalar quantum fields in finite Hilbert spaces. The main ingredients of this formalism are the dimensions $d_{\bm{k}}$ of the individual mode Hilbert spaces $\mathcal{H}_{\bm{k}}$ as well as the eigenvalue spacings $\alpha_{\bm{k}}$ and $\beta_{\bm{k}}$ of the finite-dimensional conjugate field operators $\hat Q_{\bm{k}}$ and $\hat P_{\bm{k}}$ (which were defined through Equations~\ref{eq:qp_to_QP} and \ref{eq:definition_small_q}). We have proposed a simple parametric ansatz to model the dependence of $d_{\bm{k}}$ on $|\bm{k}|$, and we argued that the number of degrees of freedom present in our field changes with the Universe's size with a sub-volume scaling as long as $d_{\bm{k}}$ is a decreasing function of $|\bm{k}|$ (and can momentarily even display an area-scaling).
For our fiducial construction, we choose $\alpha_{\bm{k}}$ and $\beta_{\bm{k}}$ such that it minimizes the impact of finite-dimensionality on the ground state energy of the mode $\bm{k}$ at the moment when the mode is initialized. We furthermore show this choice is closely tied to the requirement that both conjugate operators are equally well resolved in the ground state, making it resemble the infinite-dimensional limit as closely as possible at the time of initialization.
We have then devised accurate and numerically feasible formulae of that vacuum energy as a function of $|\bm{k}|$, which approximate the exact calculations of \cite{Singh2018}. With these formulae we were able to study how the vacuum energy density of our finite-dimensional scalar field depends of the dimensionality function $d_{\bm{k}}$. For our fiducial choice of $\alpha_{\bm{k}}$ and $\beta_{\bm{k}}$ - which minimizes finite-dimensional effects - it is decaying between two constant epochs with an overall suppression of vacuum energy by about $40\%$. And in an alternative construction, the equation of state parameter can even become $> 0$, hence causing a rapid decay that easily suppresses vacuum energy density by $\sim 10^{-60}$ compared to the infinite dimensional result (with sharp UV cut-off) for some parameter values. We have implemented the above framework within the \verb|GPUniverse| toolkit that is publicly available at \url{https://github.com/OliverFHD/GPUniverse}.

The finite-dimensional construction we have presented in this paper depends on the following non-trivial assumptions and modelling choices:
\begin{itemize}
    \item We have quantized our field in a finite box, whereas any meaningful boundary of the observable Universe would be expected to be close to spherical (\cf the discussion in \secref{size_of_cosmos}). Furthermore, our derivations have assumed that there is no spatial curvature.
    \item We have assumed that the Universe is of a constant co-moving size and we have chosen its radius to be the asymptotic future particle horizon (which has a finite co-moving size in a dark energy dominated universe).
    \item We have assumed that at any time only wave modes below a constant physical scale $\Lambda_{\mathrm{UV}}$ contribute to the field. This may be scrutinized both because we assume a sharp cut-off and because we take that cut-off to be constant in time.
    \item We made the assumption that it is the co-moving modes of the field that should be replaced by finite-dimensional operators. This, together with our assumption of a constant co-moving size of the Universe, means that the spacing of our grid in $\bm{k}$-space is constant in time. As a result, our factorisation of Hilbert space into mode sub-spaces $\mathcal{H}_{\bm{k}}$ is constant in time. This is part of a general theme of our construction: we tried to decompose our field into algebraic structures that stay constant in time.
    \item We chose to base our construction on generalised Pauli operators. These are able to mimic the concept of conjugate operator pairs, which according to \cite{Carroll_Singh2020} play a central role in the emergence of quasi-classical Hilbert space factorisations. But we have not investigated whether the GPO construction is the only way to achieve this dualism in finite dimensions, nor would we expect the emergent pointer observables of \cite{Carroll_Singh2020} to be given in terms of exact GPOs.
    \item To identify the version of the infinite-dimensional field operators which we want to replace with with GPOs, we re-aranged the scalar field Hamiltonian such that it resembles the Hamiltonian of a set of harmonic oscillators.
    \item We only considered two different choices for the spectral spacings of the field GPOs - the ones displayed in \eqnref{resolution_criterion} and \eqnref{alternative_spacing}. We tried to motivate those as representing two limiting cases: minimizing finite-dimensional effects in our fiducial choice and taking an extreme `quantum first' view in our alternative choice. But an assumption common to both constructions is that we kept the spectral spacings constant in time. If the factors of Hilbert space representing different modes $\bm{k}$ are indeed emergent and chosen such that they maximise a certain notion of classicality (as in the picture promoted by \cite{Carroll_Singh2020}) then one may speculate that the algebraic structures defined on these factors need to change with time in order to maintain that classicality.
    \item We assumed that the dimension of the mode Hilbert spaces $\mathcal{H}_{\bm{k}}$ as a function of $|\bm{k}|$ is described by \eqnref{dimension_parametrized}, \ie that it consists of a power law in $|\bm{k}|$ plus a minimum dimension $d_{\min}=2$ . This assumption has allowed us to qualitatively study consistency relations for the dimensionality of our field as well as to investigate how the dimensionality function $d_{\bm{k}}$ impacts the way in which the number of degrees of freedom in our field changes with the expansion of the Universe. But \eqnref{dimension_parametrized} is clearly an ad hoc ansatz that eventually needs to be motivated or revised by an understanding of the mapping through which spacetime and effective field theories thereon arise from an underlying quantum theory.
    \item The maximum entropy which can be attained with our construction is still much higher than would be allowed by applying the Bekenstein bound to the entire patch of the Universe we considered. According to \cite{Singh2019_essay} this may require modifying the mode density function away from the three dimensional behaviour $\dd^3k \sim k^2\dd k$ that is built into our model. 
    \end{itemize}
    Furthermore, the equation of state of our field's vacuum energy density as well as the consistency boundaries we derived on the dimensionality parameters $D$ and $n_D$ depend on a set of additional assumptions:
    \begin{itemize}
    \item We have assumed that each field mode $\bm{k}$ is initialized in its instantaneous vacuum state at the time when $\bm{k} \approx a \Lambda_{\mathrm{UV}}$ and that it evolves adiabatically after that, \ie that it remains in the (time dependent) vacuum state. This is ignoring the fact that particle production during cosmic expansion will drive our field away from its vacuum state. We leave it for further work to investigate the role of particle production, particularly in a finite-dimensional paradigm, in earlier epochs of cosmological evolution.
    \item While we have studied the equation of state of our field's vacuum energy during cosmic expansion, we have not considered this energy to be a source of that expansion. In particular, the energy density we obtained for our finite-dimensional field is still many orders of magnitude higher than the dark energy density that is needed to explain the observed accelerating expansion of our Universe \cite[\cf][]{Riess1998, Planck2018_short}. At the same time, it has been questioned whether quantum ground state energy indeed acts as a gravitational source \cite[\eg][]{Fulling1979, Yargic2020}.
    \item In the entire paper we have focused on the late-time Universe ($a \gtrsim 0.1$). An interesting line of future work would be to understand the role of finite-dimensional effects during inflation. As we discuss in \appref{late-time_expressions}, this would require modifications to our calculations because \eqnref{Emin_main_text} is not valid for arbitrarily small $a$.
\end{itemize}
The plethora of assumptions and limitations we have outlined above demonstrates, that our framework and the language we have devised to describe finite-dimensional fields still require further development. At the same time, we think that it can be a fruitful starting point to explore the impact of finite-dimensionality of Hilbert space on cosmological physics.

\acknowledgments
\textcopyright 2022. We are thankful to ChunJun (Charles) Cao, Sean Carroll, Steffen Hagstotz and Cora Uhlemann for helpful comments and discussions. OF gratefully acknowledges support by the Kavli Foundation and the International Newton Trust through a Newton-Kavli-Junior Fellowship, by Churchill College Cambridge through a postdoctoral By-Fellowship and by the Ludwig-Maximilians Universität through a Karl-Schwarzschild-Fellowship. AS acknowledges the generous support of the Heising-Simons Foundation. Part of the research described in this paper was carried out at the Jet Propulsion Laboratory, California Institute of Technology, under a contract with the National Aeronautics and Space Administration. We indebted to the invaluable work of the teams of the public python packages \verb|NumPy| \cite{NumPy}, \verb|SciPy| \cite{SciPy}, \verb|mpmath| \cite{mpmath} and \verb|Matplotlib| \cite{Matplotlib}.


\input{journaldef}
\bibliographystyle{plain}
\bibliography{main}

\appendix

\section{Eigenvalue spacing that maximize vacuum energy}
\label{app:minimizing_vacuum_energy}
 
Consider the Hamiltonian
\begin{equation}
    \hat H = \frac{\hat P^2}{2M} + \frac{M \Omega^2}{2} \hat Q^2\ ,
\end{equation}
where $\hat P$ and $\hat Q$ are generalised Pauli operators as in \cite{Singh2018, Singh2019_essay} (see also our \secref{finite_field_in_box}). In the eigenbasis of $\hat X$ this means that 
\begin{equation}
    \hat Q = \alpha\ \mathrm{diag} (-\ell,\ \dots\ ,\ \ell)\ \ \ ;\ \ \ \hat P = \beta\ \hat S\ \mathrm{diag} (-\ell,\ \dots\ ,\ \ell)\ \hat S^{-1} \ ,
\end{equation}
where $\hat S$ is Sylvester's matrix (which corresponds to discrete Fourier transform), $d = 2\ell + 1$ is the dimension of Hilbert space and the eigenvalue spacings $\alpha$ and $\beta$ satisfy $\alpha \beta = 2\pi/d$. Using this relation as well as the definition $\hat L \equiv \mathrm{diag} (-\ell,\ \dots\ ,\ \ell)$ the Hamiltonian becomes
\begin{equation}
    \hat H = \frac{1}{\alpha^2}\frac{(2\pi)^2}{2M d^2}\ \hat S \hat L^2 \hat S^{-1} + \alpha^2\frac{M \Omega^2}{2} \hat L^2\ .
\end{equation}
Since $\hat S$ is unitary, the matrix $\hat S \hat H \hat S^{-1}$ has the same eigenvalues as $\hat H$. Furthermore, it can be shown that $\hat S^2 \hat L^2 \hat S^{-2} = \hat L^2$ \cite{Singh2018}. From this it follows that the eigenspectrum of the Hamiltonian is invariant under the replacement
\begin{equation}
    \alpha \rightarrow \tilde \alpha\ \ \mathrm{with}\ \ \tilde\alpha^2\frac{M \Omega^2}{2} = \frac{1}{\alpha^2}\frac{(2\pi)^2}{2M d^2}\ .
\end{equation}
This transformation has a fixpoint for which $\alpha = \tilde \alpha$ which is given by
\begin{equation}
    \alpha_{\mathrm{fix}} = \sqrt{\frac{2\pi}{M \Omega d}}\ \ \Rightarrow\ \beta_{\mathrm{fix}} = \sqrt{\frac{2\pi M \Omega }{d}}\ \ .
\end{equation}
Because of the spectral symmetry \wrt the transformation $\alpha \rightarrow \tilde \alpha$ this fix point must extremize the minimum eigenvalue of $\hat H$. Since finite-dimensionality can only decrease the minimum eigenvalue of the Hamiltonian compared to the vacuum energy of an infinite-dimensional harmonic oscillator (\cf \cite{Singh2018} or our \appref{Emin_estimate}), it is reasonable to assume that $\alpha_{\mathrm{fix}}$ and $\beta_{\mathrm{fix}}$ indeed maximise the ground state energy of $\hat H$. Hence, they would minimize finite-dimensional effects on the low-energy spectrum of the Hamiltonian. We were not able to strictly prove the nature of the extremum, but a range of numerical tests support our assumption. These tests also show that even for low dimensions $d$ the extremum of vacuum energy lies very close to its infinite-dimensional value $\Omega / 2$.

\section{Approximating $E_{\min}(k)$}
\label{app:Emin_estimate}
\begin{figure}
\centering
  \includegraphics[width=0.7\textwidth]{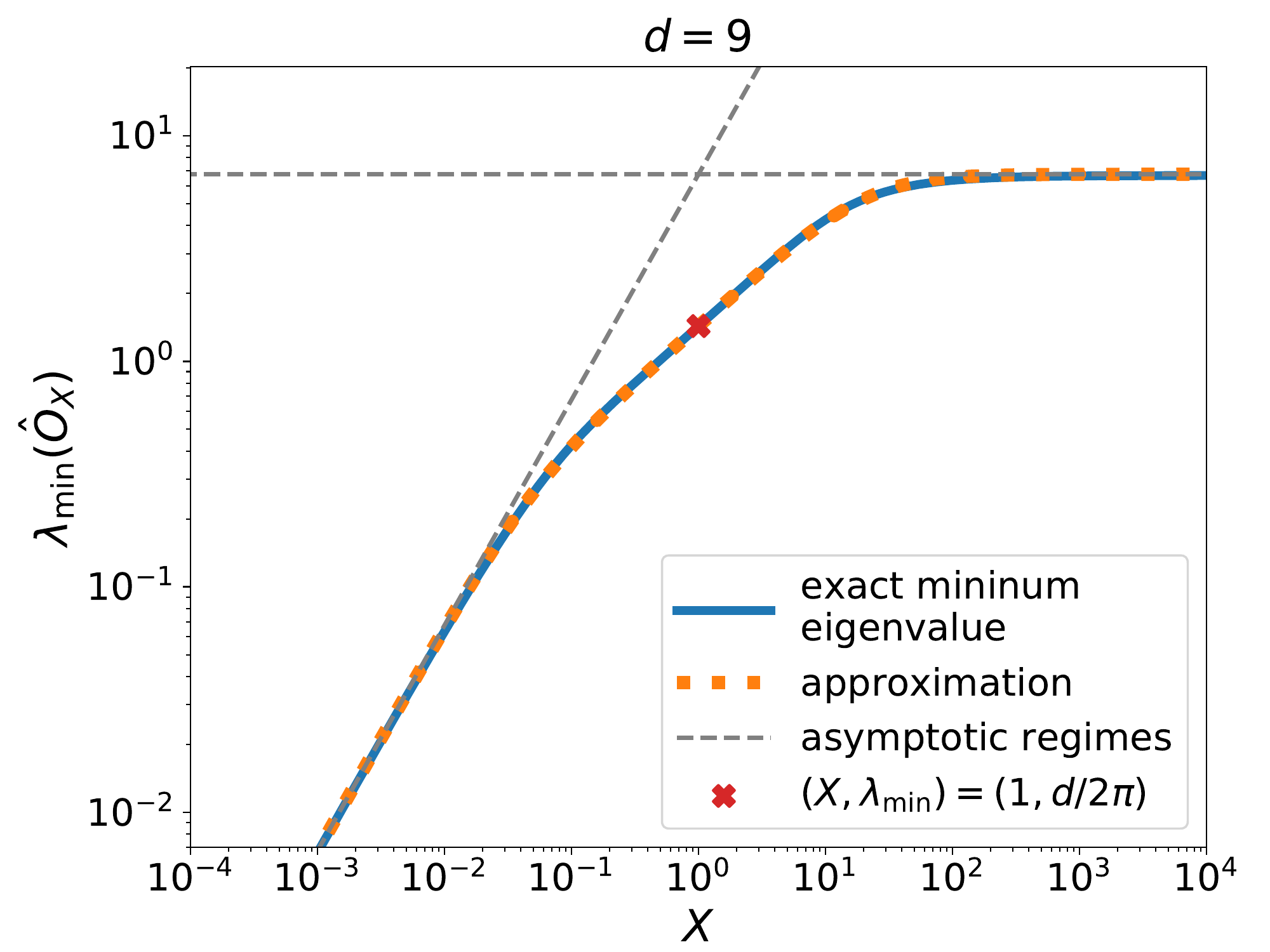}
  \vspace{0.5cm}
  
  \includegraphics[width=0.7\textwidth]{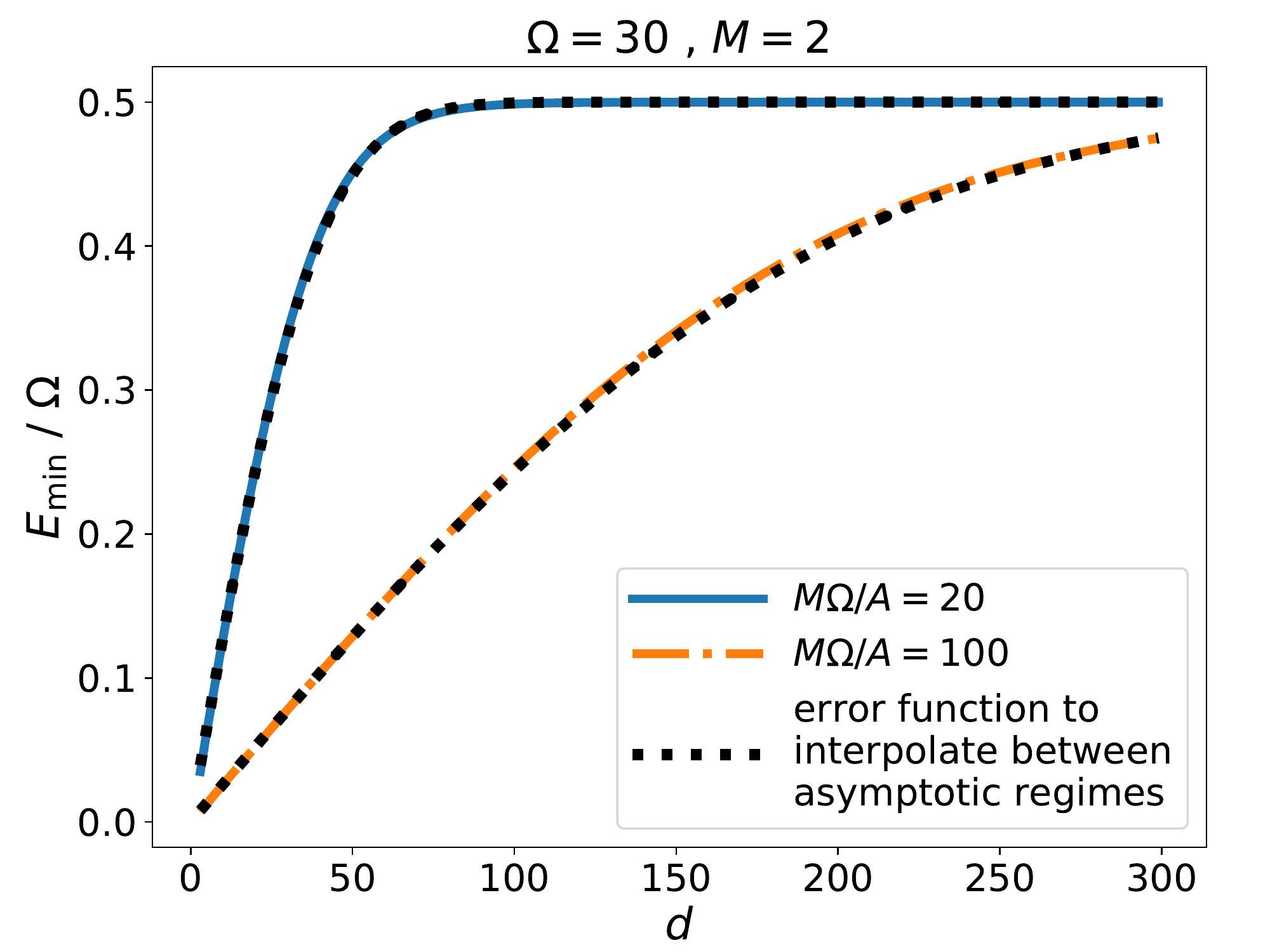}
   \caption{\textbf{Upper panel:} minimum eigenvalue of the operator $\hat O_X = \hat S \hat L^2 \hat S^{-1} + X \hat L^2$ in $d=21$ dimensions as a function of $X$. The blue solid line represents an exact calculation while the orange dotted line shows the approximation provided in \eqnref{approx_to_lambda_min_of_O_X}. The grey dashed lines and the red cross indicate the three asymptotic regimes of \eqnref{O_X_asymtotics}. \textbf{Lower panel:} minimum eigenvalue of the Hamiltonian of a finite-dimensional harmonic oscillator with (dimensionless) frequency $\Omega=30$ and mass $M=2$ as a function of Hilbert space dimension $d$ and for two different values of the parameter A (which determines how the eigenvalue spacing $\alpha$ of the position operator $\hat Q$ changes as a function of $d$, \cf \eqnref{alpha_for_cases_1_and_2}). The blue solid line represents the exact calculation for $A = M\Omega/20$ and the orange dash-dotted line shows the exact calculation for $A = M\Omega/100$. The black dotted lines show the approximation of \eqnref{approx_to_Emin}. The agreement between approximation and exact calculation is similar over a wide range of values for $\Omega$, $M$ and $A$.}
  \label{fi:lambda_min}
\end{figure}

\subsection{General case asymptotics}

Consider again the Hamiltonian of a finite-dimensional harmonic oscillator,
\begin{equation}
    \hat H = \frac{\hat P^2}{2M} + \frac{M \Omega^2}{2} \hat Q^2\ .
\end{equation}
Our goal in this appendix is to derive an approximation to the minimum eigenvalue of this operator as a function of $M$, $\Omega$, $d$ and $\alpha$ (\cf \appref{minimizing_vacuum_energy} for notation) that is numerically feasible even for large $d$. We had seen in \appref{minimizing_vacuum_energy}, that $\hat H$ can be re-written as
\begin{equation}
    \hat H = \frac{(2\pi)^2}{2 M d^2 \alpha^2}\left\lbrace \hat S \hat L^2 \hat S^{-1} + \frac{\alpha^4 M^2 \Omega^2 d^2}{(2\pi)^2} \hat L^2 \right\rbrace \equiv \frac{(2\pi)^2}{2 M d^2 \alpha^2}\left\lbrace \hat S \hat L^2 \hat S^{-1} + X \hat L^2 \right\rbrace\ .
\end{equation}
So to calculate the lowest energy level of $\hat H$ we need to know the minimum eigenvalues of operators of the form $\hat O_X \equiv \hat S \hat L^2 \hat S^{-1} + X \hat L^2$. At the fix point $\alpha_*$ we have derived in \appref{minimizing_vacuum_energy} the Hamiltonian becomes
\begin{equation}
    \hat H = \frac{\Omega}{2}\frac{2\pi}{d}\left\lbrace \hat S \hat L^2 \hat S^{-1} + \hat L^2 \right\rbrace\ .
\end{equation}
At the same time, we had argued there that the fix point minimizes effects of finite-dimensionality, such that vacuum energy comes close to its infinite-dimensional limit $\Omega/2$ (numerical calculation confirms that this is true to high accuracy even for $d$ as low as $5$). From that we can conclude that
\begin{equation}
    \lambda_{\min}(\hat O_{X\approx 1}) \approx \frac{d}{2\pi}\ .
\end{equation}
To understand the situation for more general values of $X$, let us consider the matrix elements of $\hat O_X$ in the eigenbasis of $\hat Q$. They are given by \cite{Singh2018}
\begin{align}
    [\hat O_X]_{ij} =&\ \frac{1}{4}\sum_{n\neq i,j} \frac{1}{\sin\left(\frac{2\pi \ell}{2\ell+1} (n-i) \right)\sin\left(\frac{2\pi \ell}{2\ell + 1} (n-j) \right)} + X j^2 \delta_{ij}\ ,
\end{align}
where all integers run from $-\ell$ to $\ell$ (and $d = 2\ell +1$ as in \appref{minimizing_vacuum_energy}). In the limit $X \rightarrow \infty$ the second term in the above bracket dominates such that the eigenvector $\bm{v}$ of $\hat O_X$ with the lowest eigenvalue becomes
\begin{equation}
    [\bm{v}]_i \propto \left\lbrace\begin{matrix}
    1 &\ \mathrm{for}\ i=0 \\
    0 &\ \mathrm{else}
    \end{matrix}\right.
\end{equation}
with the corresponding eigenvalue
\begin{equation}
    \lambda_{\min}(\hat O_X) \approx [\hat O_X]_{00} = \frac{1}{2}\sum_{n=1}^\ell \frac{1}{\sin^2\left(\frac{2\pi \ell}{2\ell+1} n \right)} = \frac{1}{2}\sum_{n=1}^\ell \frac{1}{\sin^2\left(\pi n (1 - 1/d) \right)}\ .
\end{equation}
For $d \gg 1$ we can expand this in the small parameter $x \equiv 1/d$ as
\begin{equation}
\label{eq:OX00_approx}
    [\hat O_X]_{00} \approx \frac{1}{2 \pi^2 x^2} \sum_{n=1}^\ell \frac{1}{n^2} \approx \frac{d^2}{12}\ .
\end{equation}
This is an approximation for $\lambda_{\min}(\hat O_X)$ as $X$ approaches infinity. By the symmetry arguments we had employed in \appref{minimizing_vacuum_energy} one can then show that $\lambda_{\min}(\hat O_X) \approx X d^2/12$ as $X$ approaches zero. In summary, we obtain
\begin{equation}
\label{eq:O_X_asymtotics}
    \lambda_{\min}(\hat O_X) \approx \left\lbrace\begin{matrix}
    d^2/12 &\ \mathrm{as}\ X \rightarrow \infty\\
    \\
    d / (2\pi) &\ \mathrm{for}\ X \approx 1\\
    \\
    X d^2 / 12 &\ \mathrm{as}\ X \rightarrow 0
    \end{matrix}\right.\ .
\end{equation}
These asymptotics are matched exactly by the formula
\begin{equation}
    \label{eq:approx_to_lambda_min_of_O_X}
    \lambda_{\min}(\hat O_X) \approx \frac{F}{\left(1 + X^{-3} + G X^{-3/2} \right)^{1/3}}
\end{equation}
with
\begin{equation}
    F = \frac{d^2}{12}\ \ ,\ \ G = \left(\frac{\pi d}{6}\right)^3 - 2\ .
\end{equation}
We compare that formula to the exact calculation of $\lambda_{\min}(\hat O_X)$ in the upper panel of \figref{lambda_min} - exact and approximated result agree to within $\sim 4\%$ accuracy over the entire range of $X$ for $d\geq 7$ (and much better for most values of $X$). That accuracy reduces to $\sim 6\%$ for $d = 5$ and to $\sim 15\%$ for $d = 3$.

\subsection{Expressions for late-time cosmology}
\label{app:late-time_expressions}

We can obtain a more concise formula that directly approximates the minimum eigenvalue of $\hat H$ for the late-time cosmological situation of \secref{finite_dim_eos}. For both constructions we considered there, the eigenvalues spacing of $\hat Q$ was of the form
\begin{equation}
\label{eq:alpha_for_cases_1_and_2}
    \alpha = \sqrt{\frac{2\pi}{Ad}}\ ,
\end{equation}
with $A = M(t_{\bm{k}}) \Omega_{\bm{k}}(t_{\bm{k}})$ in our fiducial construction and $A = 1$ in \secref{alternative_spacing}. The parameter $X$ then becomes $X = M^2 \Omega^2 / A^2$,
which in our fiducial construction is given as a function of $\bm{k}$ and $t$ by
\begin{equation}
    X_{\bm{k}}^{\mathrm{fid}}(t) = \frac{M(t)^2 \Omega(t)^2}{M(t_{\bm{k}})^2 \Omega(t_{\bm{k}})^2} \approx \left(\frac{a(t)}{a(t_{\bm{k}})}\right)^4\ .
\end{equation}
In an expanding universe this is clearly always larger than $1$. On the other hand, in our alternative construction $X$ becomes
\begin{equation}
    X_{\bm{k}}^{\mathrm{alt}}(t) = L_c^2 a^4 k^2 = (L_c a^3 k_{\mathrm{ph}})^2\ .
\end{equation}
In the late-time Universe, $a\sim 1$, this is greater than $1$ as long as $k \gtrsim 1/L_c \approx 3.3\cdot 10^{-34}$ eV, \ie on all scales relevant to the vacuum energy density of our scalar field. So for the late-time expansion we have considered in this work, we are indeed fine to consider only the two asymptotics of \eqnref{O_X_asymtotics} with $X \geq 1$.
In that case, and taking into account the relation between $\hat H$ and $\hat O_X$, the minimum energy eigenvalue will asymptotically behave as
\begin{align}
    \lambda_{\min}(\hat H) \approx&\ \frac{A}{2M} \cdot\left\lbrace\begin{matrix}
    \pi d/6 &\ \mathrm{as}\ M \Omega/A \rightarrow \infty \\
    \\
    1 &\ \mathrm{for}\ M \Omega/A \approx 1
    \end{matrix}\right.\nonumber \\
    \approx&\ \frac{\Omega}{2} \cdot\left\lbrace\begin{matrix}
    \frac{\pi A d}{6 M \Omega} &\ \mathrm{as}\ M \Omega/A \rightarrow \infty \\
    \\
    1 &\ \mathrm{for}\ M \Omega/A \approx 1
    \end{matrix}\right.\ .
\end{align}
This behaviour can be approximately matched by the ansatz
\begin{equation}
    \label{eq:approx_to_Emin}
    \lambda_{\min}(\hat H) \approx \frac{\Omega}{2} \erf\left( \frac{\pi^{3/2} A d}{12 M \Omega} \right)\ .
\end{equation}
This is the approximation we used in order to derive the results of \secref{finite_dim_eos}. We compare it to the exact calculation of $\lambda_{\min}(\hat H)$ over a limited range of $d$ in the lower panel of \figref{lambda_min}. That figure only includes results for one set of values for $\Omega$ and $M$ and for two different values for $A$. But we find that \eqnref{approx_to_Emin} agrees with the exact result to within $\sim 2\%$ accuracy for a wide range of values for $(\Omega, M, A)$ as long as $d \geq 7$. For  $d = 5$ this reduces to $\sim 4\%$ accuracy and for $d = 3$ to $\sim 12\%$ accuracy.

Finally, we want to note that in the early Universe the third regime of \eqnref{O_X_asymtotics} can indeed become relevant - at least in our alternative construction. There one would have $X < 1$ even at $k_{\mathrm{ph}} \approx \Lambda_{\mathrm{UV}} = 1$ as long as $a \lesssim 3.0\cdot 10^{-21}$. So the approximation of \eqnref{approx_to_Emin} would \eg be not appropriate to calculate the behaviour of our finite-dimensional field during inflation, and one would have to use \eqnref{approx_to_lambda_min_of_O_X} instead. Both \eqnref{approx_to_Emin}, \eqnref{approx_to_lambda_min_of_O_X} and the corresponding exact calculations are all implemented within our publicly available code package \verb|GPUniverse|.

\section{Error function asymptotics, alternative construction}
\label{app:erf_asymptotics}

\begin{figure}
\centering
  \includegraphics[width=0.9\textwidth]{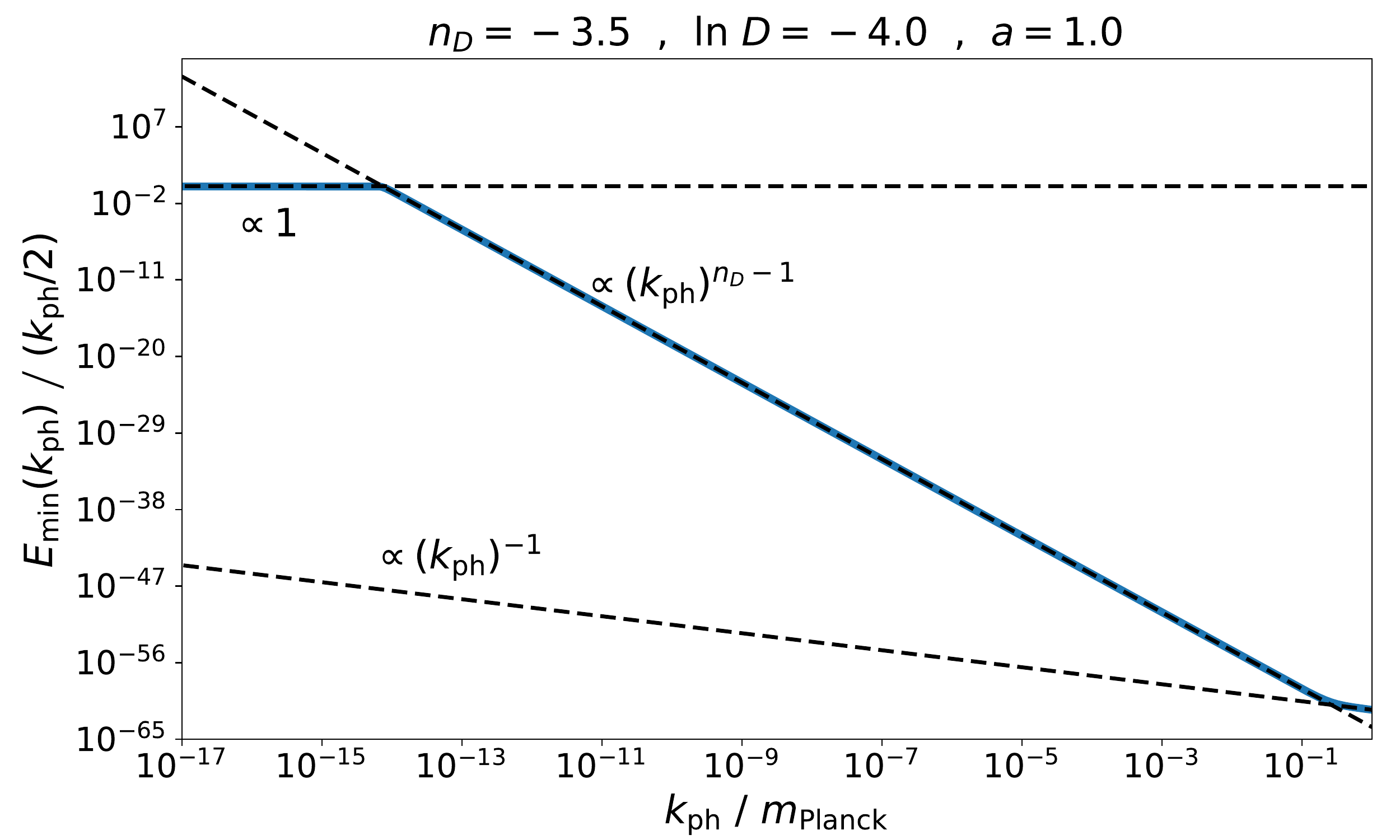}
   \caption{Displaying the upper panel of \figref{integrands} for $a=1$ on a wider range of $k_{\mathrm{ph}}$ (solid blue line) and indicating the asymptotic behaviour stated by \eqnref{alternative_asymtotics} (black dashed lines). With the alternative eigenvalue spacings $\alpha_{\bm{k}}$ and  $\beta_{\bm{k}}$ we studied in \secref{alternative_spacing} (\cf \eqnref{alternative_spacing}) the vacuum energy in each co-moving mode $\bm{k} = a \bm{k}_{\mathrm{ph}}$ can strongly deviate from the infinite-dimensional limit $E_{\min} = k_{\mathrm{ph}}/2$. This is in contrast to our fiducial construction where for the same parameters these deviations stay moderate even at high energy scales (\cf lower panel of \figref{integrands}).}
  \label{fi:Integrant_alternative_case_asymptotes}
\end{figure}

For $a=1$ we repeat the upper panel of \figref{integrands} on a wider range of $k_{\mathrm{ph}}$ in \figref{Integrant_alternative_case_asymptotes} (solid blue line in that figure). We also show that \eqnref{alternative_asymtotics} is indeed an accurate description of the asymptotic behaviour of the error function appearing in \eqnref{eps_vac_alternative_spacings} (\cf the black dashed line in \figref{Integrant_alternative_case_asymptotes}).

\end{document}

%% file: journaldef.tex
\def\aj{AJ}%
\def\araa{ARA\&A}%
\def\apj{ApJ}%
\def\apjl{ApJ}%
\def\apjs{ApJS}%
\def\ao{Appl.~Opt.}%
\def\apss{Ap\&SS}%
\def\aap{A\&A}%
\def\aapr{A\&A~Rev.}%
\def\aaps{A\&AS}%
\def\azh{AZh}%
\def\baas{BAAS}%
\def\jrasc{JRASC}%
\def\memras{MmRAS}%
\def\mnras{MNRAS}%
\def\pra{Phys.~Rev.~A}%
\def\prb{Phys.~Rev.~B}%
\def\prc{Phys.~Rev.~C}%
\def\prd{Phys.~Rev.~D}%
\def\pre{Phys.~Rev.~E}%
\def\prl{Phys.~Rev.~Lett.}%
\def\pasp{PASP}%
\def\pasj{PASJ}%
\def\qjras{QJRAS}%
\def\skytel{S\&T}%
\def\solphys{Sol.~Phys.}%
\def\sovast{Soviet~Ast.}%
\def\ssr{Space~Sci.~Rev.}%
\def\zap{ZAp}%
\def\nat{Nature}%
\def\iaucirc{IAU~Circ.}%
\def\aplett{Astrophys.~Lett.}%
\def\apspr{Astrophys.~Space~Phys.~Res.}%
\def\bain{Bull.~Astron.~Inst.~Netherlands}%
\def\fcp{Fund.~Cosmic~Phys.}%
\def\gca{Geochim.~Cosmochim.~Acta}%
\def\grl{Geophys.~Res.~Lett.}%
\def\jcap{JCAP}%
\def\jcp{J.~Chem.~Phys.}%
\def\jgr{J.~Geophys.~Res.}%
\def\jqsrt{J.~Quant.~Spec.~Radiat.~Transf.}%
\def\memsai{Mem.~Soc.~Astron.~Italiana}%
\def\nphysa{Nucl.~Phys.~A}%
\def\physrep{Phys.~Rep.}%
\def\physscr{Phys.~Scr}%
\def\planss{Planet.~Space~Sci.}%
\def\procspie{Proc.~SPIE}%